\documentclass[
 reprint,
 superscriptaddress,
nofootinbib,
 amsmath,amssymb,
 aps,
 prd,
floatfix,
]{revtex4-2}
\usepackage[colorlinks=true,linkcolor=blue,citecolor=blue,urlcolor=blue]{hyperref}
\usepackage[separate-uncertainty,retain-explicit-plus,per-mode=symbol,binary-units,range-phrase=--,range-units=single]{siunitx}[=v2]
\usepackage[export]{adjustbox}
\usepackage{wrapfig}
\usepackage{ragged2e}
\usepackage{array,mathtools,dcolumn}
\newcolumntype{x}[1]{>{\centering\arraybackslash\hspace{0pt}}p{#1}}
\usepackage{amsmath,amssymb}
\usepackage{wasysym}
\usepackage{stmaryrd}
\usepackage[below]{placeins}
\usepackage[table,xcdraw]{xcolor}
\usepackage{soul}
\usepackage{tikz}
\usepackage{afterpage}
\usepackage{lineno}
\usepackage{listings}
\usepackage{array}
 \usepackage[normalem]{ulem}
 \useunder{\uline}{\ul}{}
\makeatletter
\@ifundefined{c@rownum}{
  \let\c@rownum\rownum
}{}
\@ifundefined{therownum}{
  \def\therownum{\@arabic\rownum}
}{}
\makeatother
	\newcounter{rowno}
\usepackage[version=4]{mhchem}
\usepackage{multirow}
\usepackage{eurosym}
\usepackage{pagecolor}
\usepackage{fancyhdr}
\usepackage{etoolbox}
\usepackage{calc}
\usepackage{dcolumn}

\usepackage{bm}
\usepackage{xargs}
\usepackage{xfrac}
\newcommandx{\unsure}[2][1=]{\todo[linecolor=red,backgroundcolor=red!25,bordercolor=red,#1]{#2}}
\newcommandx{\change}[2][1=]{\todo[linecolor=blue,backgroundcolor=blue!25,bordercolor=blue,#1]{#2}}
\newcommandx{\info}[2][1=]{\todo[linecolor=OliveGreen,backgroundcolor=OliveGreen!25,bordercolor=OliveGreen,#1]{#2}}
\newcommandx{\improvement}[2][1=]{\todo[linecolor=Plum,backgroundcolor=Plum!25,bordercolor=Plum,#1]{#2}}
\newcommandx{\thiswillnotshow}[2][1=]{\todo[disable,#1]{#2}}

\newcommand{\refcite}[1]{Ref.~\cite{#1}}
\newcommand{\refscite}[1]{Refs.~\cite{#1}}
\newcommand{\reffig}[1]{Fig.~\ref{#1}}
\newcommand{\reftab}[1]{Table~\ref{#1}}
\newcommand{\refeqn}[1]{Eq.~(\ref{#1})}
\newcommand{\refsec}[1]{Sec.~\ref{#1}}

\usepackage{paralist}
\usepackage{tcolorbox}
\usepackage{pgfornament}
\usepackage{longtable}
\usepackage{hhline}
\usepackage{array}
\usepackage[nodayofweek]{datetime}
\usepackage{subfigure}
\usepackage{import}
\usepackage{appendix}
\usepackage{cleveref}
\usepackage{ellipsis}
\usepackage{enumitem}
\setdescription{itemsep=0pt,parsep=0pt,labelsep=.2cm,leftmargin=2.2cm,labelwidth=2cm,listparindent=.7cm}
\setlist{nolistsep,leftmargin=1cm}
\newlist{enumcompactitem}{itemize}{3}
\setlist[enumcompactitem]{topsep=0pt,partopsep=0pt,itemsep=0pt,parsep=0pt}
\setlist[enumcompactitem,1]{label=\textbullet}
\setlist[enumcompactitem,2]{label=---}
\setlist[enumcompactitem,3]{label=*}
\newlist{enumcompactdesc}{description}{3}
\setlist[enumcompactdesc]{topsep=0pt,partopsep=0pt,itemsep=0pt,parsep=0pt}
\newlist{enumcompactenum}{enumerate}{3}
\setlist[enumcompactenum]{topsep=0pt,partopsep=0pt,itemsep=0pt,parsep=0pt}
\setlist[enumcompactenum,1]{label=\arabic*}
\setlist[enumcompactenum,2]{label=\alph*}
\setlist[enumcompactenum,3]{label=\roman*}
\newcommand{\etal}{\textit{et al.}}

\newcommand{\GFDS}{\mbox{\tt G4DS}}

\newcommand{\Geant}{\mbox{\tt Geant4}}
\newcommand{\FLUKA}{\mbox{\tt FLUKA}}

\newcommand{\EXFOR}{\mbox{\tt EXFOR}}
\newcommand{\COSMO}{\mbox{\tt COSMO}}
\newcommand{\YIELDX}{\mbox{\tt YIELDX}}
\newcommand{\ACTIVIA}{\mbox{\tt ACTIVIA}}
\newcommand{\TENDL}{\mbox{\tt TENDL}}
\newcommand{\HEAD}{\mbox{\tt HEAD-2009}}
\newcommand{\JENDL}{\mbox{\tt JENDL/AN-2005}}
\newcommand{\NEST}{\mbox{\tt NEST}}

\newcommand{\isotope}[2]{\mbox{$^{#1}$#2}}
\newcommand{\exsitu}{\emph{ex situ}}
\newcommand{\insitu}{\emph{in situ}}
\newcommand{\SE}{\mbox{SE}} 
\newcommand{\SEs}{\mbox{SEs}}

\newcommand{\eg}{\mbox{\emph{e.g.}}}

\DeclareSIUnit\c{\mbox{$c$}}
\DeclareSIUnit\magn{\mbox{$\times$}}
\DeclareSIUnit\min{min}
\DeclareSIUnit\hr{hr}
\DeclareSIUnit\hrs{hrs}
\DeclareSIUnit\week{week}
\DeclareSIUnit\month{mo}
\DeclareSIUnit\months{mos}
\DeclareSIUnit\year{yr}
\DeclareSIUnit\years{years}
\DeclareSIUnit\yr{yr}
\DeclareSIUnit\standard{std}
\DeclareSIUnit\str{sr}
\DeclareSIUnit\ppm{ppm}
\DeclareSIUnit\ppb{ppb}
\DeclareSIUnit\ppt{ppt}
\DeclareSIUnit\pe{PE}
\DeclareSIUnit\spe{SPE}
\DeclareSIUnit\pdm{PDM}
\DeclareSIUnit\ev{events}
\DeclareSIUnit\ct{counts}
\DeclareSIUnit\neutron{\mbox{$n$}}
\DeclareSIUnit\smp{samples}
\DeclareSIUnit\Sample{S}
\DeclareSIUnit\ch{ch}
\DeclareSIUnit\hit{hit}
\DeclareSIUnit\hits{hits}
\DeclareSIUnit\bin{(\mbox{5-PE}~bin)}
\DeclareSIUnit\sgm{\mbox{$\sigma$}}
\DeclareSIUnit\rms{RMS}
\DeclareSIUnit\keVee{\mbox{keV$_{{\rm ee}}$}}
\DeclareSIUnit\keVr{\mbox{keV$_{\rm nr}$}}
\DeclareSIUnit\eVee{\mbox{eV$_{\rm ee}$}}
\DeclareSIUnit\eVr{\mbox{eV$_{\rm nr}$}}
\DeclareSIUnit\ph{photon}
\DeclareSIUnit\el{\mbox{$e^-$}}
\DeclareSIUnit\pm{\mbox{PMT}}
\DeclareSIUnit\pixel{\mbox{pixel}}
\DeclareSIUnit\inch{''}
\DeclareSIUnit\foot{'}
\DeclareSIUnit\bit{bit}
\DeclareSIUnit\sample{samples}
\DeclareSIUnit\barn{barn}
\DeclareSIUnit\bara{bar}
\DeclareSIUnit\bar{bar}
\DeclareSIUnit\barg{barg}
\DeclareSIUnit\mlardepth{\mbox(meter~of~\LAr~depth)}
\DeclareSIUnit\Curie{Ci}
\DeclareSIUnit\psia{psia}
\DeclareSIUnit\psf{psf}
\DeclareSIUnit\pcf{pcf}
\DeclareSIUnit\parsec{pc}
\DeclareSIUnit\cps{cps}
\DeclareSIUnit\mwe{\mbox{m.w.e.}}
\DeclareSIUnit\liveday{\mbox{live-days}}
\DeclareSIUnit\days{\mbox{days}}
\DeclareSIUnit\miles{\mbox{miles}}
\DeclareSIUnit\lumens{\mbox{lm}}
\DeclareSIUnit\degreeC{\mbox{$^{\circ}$C}}
\DeclareSIUnit\degreeF{\mbox{$^{\circ}$F}}
\DeclareSIUnit\electron{\mbox{$e^-$}}
\DeclareSIUnit\Euro{\mbox{\euro}}
\DeclareSIUnit\cph{cph}
\DeclareSIUnit\neq{neq}
\DeclareSIUnit\normal{\mbox{N}}
\DeclareSIUnit\USD{\mbox{\$}}

\newcommand{\PbTwoOneZeroHalfLife}{\SI{22.3}{\yr}}

\newcommand{\alphan}{($\alpha$,$n$)}
\newcommand{\DM}{\mbox{DM}}

\newcommand{\CEnNS}{\mbox{CE$\nu$NS}}
\newcommand{\DS}{\mbox{DarkSide}}

\newcommand{\DSf}{\mbox{DarkSide-50}}
\newcommand{\DSk}{\mbox{DarkSide-20k}}
\newcommand{\DSl}{\mbox{DarkSide-LowMass}}

\newcommand{\DEAP}{\mbox{DEAP-3600}}
\newcommand{\SCENE}{\mbox{SCENE}}
\newcommand{\ARIS}{\mbox{ARIS}}
\newcommand{\Urania}{\mbox{Urania}}
\newcommand{\Aria}{\mbox{Aria}}
\newcommand{\TPC}{\mbox{TPC}}
\newcommand{\TPCs}{\mbox{\TPC s}}
\newcommand{\LArTPC}{\mbox{\LAr~\TPC}}
\newcommand{\LArTPCs}{\mbox{\LAr~\TPCs}}

\newcommand{\SiPMs}{\mbox{SiPMs}}
\newcommand{\PDE}{\mbox{PDE}}
\newcommand{\WIMP}{\mbox{WIMP}}
\newcommand{\WIMPs}{\mbox{\WIMP s}}

\newcommand{\TPB}{\mbox{TPB}}

\newcommand{\LAr}{\ce{LAr}}

\newcommand{\UAr}{\ce{UAr}}
\newcommand{\LNGS}{\mbox{LNGS}}
\newcommand{\LNGSfullname}{\textit{Laboratori Nazionali del Gran Sasso}}
\newcommand{\BUL}{Boulby Underground Laboratory}
\newcommand{\SNOLAB}{\mbox{SNOLAB}}
\newcommand{\XENONOT}{\mbox{XENON1T}}

\newcommand{\Clevios}{\mbox{Clevios\textsuperscript{TM}}}

\newcommand{\DarkPhotonMassSymbol}{\mbox{m$_{A^\prime}$}}
\newcommand{\DarkPhotonAlphaSymbol}{\mbox{$\alpha_D$}}
\newcommand{\WIMPMassSymbol}{\mbox{m$_\chi$}}
\newcommand{\WIMPMassLowLimit}{\SI{10}{\GeV\per\square\c}}

\newcommand{\WIMPMassTenTev}{\SI{10}{\TeV\per\square\c}}
\newcommand{\NinetyPerCentCL}{\mbox{$90\%$~C.L.}}

\newcommand{\NR}{\mbox{NR}}

\newcommand{\ER}{\mbox{ER}}
\newcommand{\ERs}{\mbox{ERs}}

\newcommand{\gr}{\mbox{$\gamma$-ray}}
\newcommand{\grs}{\mbox{$\gamma$-rays}}
												
\newcommand{\xr}{\mbox{X-ray}}
\newcommand{\xrs}{\mbox{X-rays}}
\newcommand{\bta}{\mbox{$\beta$}}

\newcommand{\SOne}{\mbox{S1}}
\newcommand{\STwo}{\mbox{S2}}
\newcommand{\gTwo}{\mbox{$g_2$}}

\newcommand{\Ne}{\mbox{$N_{e^-}$}}

\newcommand{\DSlSurfaceArea}{\SI{5.7}{\square\meter}}

\newcommand{\DSLowMassMidMass}{\mbox{\SI{5}{\GeV\per\square\c}}}
\newcommand{\PDM}{\mbox{PDM}}
\newcommand{\PDMs}{\mbox{\PDM s}}

\newcommand{\AriaSeruciOneRate}{\SI{8(2)}{\kg\per\day}}
\newcommand{\AriaDepletionPerPass}{\num{10}}

\newcommand{\DSfUArArThreeNineActivity}{\SI{0.73(11)}{\milli\becquerel\per\kg}}	
\newcommand{\DSfUArArThreeNineActivityValMicro}{\num{730(110)}}

\newcommand{\DSfLArBelowMeshElectronSpeedNum}{\num{0.93}}				
\newcommand{\DSfDdKrEightFiveActivityNumMicro}{\num{1900\pm100}}	
							
\newcommand{\DSfDdKrEightFiveActivity}{\SI{1.9\pm0.1}{\milli\becquerel\per\kg}}
\newcommand{\CollectionEfficiencySymbol}{\mbox{$\epsilon_\text{ph}$}}
\newcommand{\STwoPhotonYieldSymbol}{\mbox{$Y_\text{ph}^\text{S2}$}}
\newcommand{\DSLMBathVetoUArMass}{\SI{4.5}{\tonne}}
\newcommand{\DSLMBathVetoThreshold}{\SI{100}{\keV}}
\newcommand{\DSLMPDMBufferVetoThreshold}{\SI{50}{\keV}}
\newcommand{\DSLMBathVetoThickness}{\SI{28}{\cm}}
\newcommand{\DSLMCryostatWallThickness}{\SI{0.5}{\cm}}
\newcommand{\DSLMCryostatHeight}{\SI{215}{\cm}}
\newcommand{\DSLMCryostatDiameter}{\SI{170}{\cm}}

\newcommand{\DSLMGasPocketThickness}{\SI{1}{\cm}}
\newcommand{\DSLMTPCHeight}{\SI{111}{\cm}}
\newcommand{\DSLMTPCDiameter}{\SI{110}{\cm}}
\newcommand{\DSLMXYResolutionNum}{2.8}
\newcommand{\DSLMXYResolution}{\SI{\DSLMXYResolutionNum}{\cm}}
\newcommand{\DSLMCollectionEfficiencyNum}{0.27}
\newcommand{\DSLMCollectionEfficiency}{\SI{\DSLMCollectionEfficiencyNum}{\pe\per\ph}}
\newcommand{\DSLMSTwoPhotonYieldNum}{280}
\newcommand{\DSLMSTwoPhotonYield}{\SI{\DSLMSTwoPhotonYieldNum}{\ph\per\el}}
\newcommand{\DriftFieldSymbol}{\mbox{$\mathcal{E}_d$}}
\newcommand{\DriftFieldSymbolPow}{\mbox{$\mathcal{E}_d^B$}}
\newcommand{\tdrift}{\mbox{$t_{\text{drift}}$}}
\newcommand{\tdriftmax}{\mbox{$t_{\text{drift}}^{\text{max}}$}}
\newcommand{\vdrift}{\mbox{$v_{\rm drift}$}}

\newcommand{\cBoxER}{\mbox{$c_\Box^\text{ER}$}}
\newcommand{\cBoxNR}{\mbox{$c_\Box^\text{NR}$}}
\newcommand{\fB}{\mbox{$f_B$}}
\newcommand{\EFieldPower}{\mbox{$B$}}
\newcommand{\StoppingPowerElec}{\mbox{$S_e$}}
\newcommand{\StoppingPowerNucl}{\mbox{$S_n$}}
\newcommand{\NeER}{\mbox{$N_e^\text{ER}$}}
\newcommand{\NeNR}{\mbox{$N_e^\text{NR}$}}
\newcommand{\NeNRER}{\mbox{$N_e^\text{NR,ER}$}}
\newcommand{\pe}{\mbox{PE}}
\newcommand{\ChargeYieldER}{\mbox{$Q_y^{\rm ER}$}}
\newcommand{\ChargeYieldNR}{\mbox{$Q_y^{\rm NR}$}}
\newcommand{\ChargeYieldNRER}{\mbox{$Q_y^{\rm NR,ER}$}}
\newcommand{\XYResolutionSymbol}{\mbox{$\sigma_{xy}$}}
\newcommand{\DSLMVetoWindow}{\mbox{\SI{1.18}{\ms}}}
\newcommand{\DSLMPeakPDEEfficiency}{\SI{40}{\percent}}
\newcommand{\DSLMDriftField}{\SI{200}{\volt\per\cm}}
\newcommand{\DSLMDriftFieldStudyRange}{{\SIrange{200}{1000}{\volt\per\cm}}}
\newcommand{\DSLMELField}{\SI{6.5}{\kilo\volt\per\cm}}
\newcommand{\DSLMgTwoNominalNum}{75}
\newcommand{\DSLMgTwoNominal}{\mbox{\SI{\DSLMgTwoNominalNum}{\pe\per\el}}}

\newcommand{\DSLMnuESBkgd}{\SI{13.4\pm0.4}{\ev\per\tonne\per\year}} 
\newcommand{\DSLMNumPDMs}{\num{864}}

\newcommand{\DSLMTargetActiveMass}{\SI{1.5}{\tonne}}
\newcommand{\DSLMTargetFiducialMass}{\SI{1}{\tonne}}
\newcommand{\DSLMTargetFiducialRadius}{\SI{45}{\cm}}
\newcommand{\DSLMPDMOffset}{\SI{10}{\cm}}
\newcommand{\DSfUArArThreeNineActivityOverTenVal}{73}
\newcommand{\DSfUArArThreeNineActivityOverTen}{\SI{\DSfUArArThreeNineActivityOverTenVal}{\micro\becquerel\per\kg}}
\newcommand{\DSfUArArThreeNineActivityOverOneHundredVal}{7.3}
\newcommand{\DSfUArArThreeNineActivityOverOneHundred}{\SI{\DSfUArArThreeNineActivityOverOneHundredVal}{\micro\becquerel\per\kg}}
 
\newcommand{\DSLMSurfaceBkgdDepth}{\SI{50}{\micro\meter}} 
\newcommand{\DSLMExposure}{\SI{1}{\tonne\year}}
\newcommand{\AArArFourTwoActivity}{\SI{40.4\pm5.0}{\micro\becquerel\per\kg}}

\usepackage{graphicx}
\usepackage{dcolumn}
\usepackage{mhchem}
\usepackage{bm}
\usepackage{makecell}
\usepackage{titlesec}
\usepackage{cellspace} 
\usepackage{subfigure} 
\usepackage{float} 
\usepackage{pdfcomment}

\titlespacing*{\subsection}{0pt}{1\baselineskip}{1\baselineskip} 

\usepackage{graphicx}
\begin{document} 
\title{Sensitivity projections for a dual-phase argon TPC optimized for light dark matter searches through the ionization channel}
\newcommand{\Alberta}{Department of Physics, University of Alberta, Edmonton, AB T6G 2R3, Canada}
\newcommand{\APC}{APC, Universit\'e de Paris, CNRS, Astroparticule et Cosmologie, Paris F-75013, France}
\newcommand{\AQLNGS}{INFN Laboratori Nazionali del Gran Sasso, Assergi (AQ) 67100, Italy}
\newcommand{\AQGSSI}{Gran Sasso Science Institute, L'Aquila 67100, Italy}
\newcommand{\AQUni}{Università degli Studi dell'Aquila, L'Aquila 67100, Italy}
\newcommand{\AstroCeNT}{AstroCeNT, Nicolaus Copernicus Astronomical Center of the Polish Academy of Sciences, 00-614 Warsaw, Poland}
\newcommand{\Augustana}{Physics Department, Augustana University, Sioux Falls, SD 57197, USA}
\newcommand{\Belgorod}{Radiation Physics Laboratory, Belgorod National Research University, Belgorod 308007, Russia}
\newcommand{\BHSU}{School of Natural Sciences, Black Hills State University, Spearfish, SD 57799, USA}
\newcommand{\BINP}{Budker Institute of Nuclear Physics, Novosibirsk 630090, Russia}
\newcommand{\Birmingham}{School of Physics and Astronomy, University of Birmingham, Edgbaston, B15 2TT, Birmingham, UK}
\newcommand{\BNLaddress}{Brookhaven National Laboratory, Upton, NY 11973, USA}
\newcommand{\BOINFN}{INFN Bologna, Bologna 40126, Italy}
\newcommand{\BOUniPHY}{Department of Physics and Astronomy, Universit\`a degli Studi di Bologna, Bologna 40126, Italy}
\newcommand{\CAUniCHE}{Department of Mechanical, Chemical, and Materials Engineering, Universit\`a degli Studi, Cagliari 09042, Italy}
\newcommand{\CAUniEEE}{Department of Electrical and Electronic Engineering, Universit\`a degli Studi di Cagliari, Cagliari 09123, Italy}
\newcommand{\CAUniPHY}{Physics Department, Universit\`a degli Studi di Cagliari, Cagliari 09042, Italy}
\newcommand{\CAINFN}{INFN Cagliari, Cagliari 09042, Italy}
\newcommand{\Carleton}{Department of Physics, Carleton University, Ottawa, ON K1S 5B6, Canada}
\newcommand{\Campinas}{Physics Institute, Universidade Estadual de Campinas, Campinas 13083, Brazil}
\newcommand{\Columbia}{Physics Department, Columbia University, New York, NY 10027, USA}
\newcommand{\Chicago}{Department of Physics and Kavli Institute for Cosmological Physics, University of Chicago, Chicago, IL 60637, USA}
\newcommand{\CentroFermi}{Museo Storico della Fisica e Centro Studi e Ricerche Enrico Fermi, Roma 00184, Italy}
\newcommand{\CERNaddress}{CERN, European Organization for Nuclear Research 1211 Geneve 23, Switzerland, CERN}
\newcommand{\CIEMAT}{CIEMAT, Centro de Investigaciones Energ\'eticas, Medioambientales y Tecnol\'ogicas, Madrid 28040, Spain}
\newcommand{\Cluj}{National Institute for R\&D of Isotopic and Molecular Technologies, Cluj-Napoca, 400293, Romania}
\newcommand{\CPPM}{Centre de Physique des Particules de Marseille, Aix Marseille Univ, CNRS/IN2P3, CPPM, Marseille, France}
\newcommand{\CTINFN}{INFN Catania, Catania 95121, Italy}
\newcommand{\CTUNI}{Universit\`a of Catania, Catania 95124, Italy}
\newcommand{\CTLNS}{INFN Laboratori Nazionali del Sud, Catania 95123, Italy}
\newcommand{\ENSMP}{\'Ecole nationale sup\'erieure des mines de Paris, Paris 75272, France}
\newcommand{\ENUniCEE}{Engineering and Architecture Faculty, Universit\`a di Enna Kore, Enna 94100, Italy}
\newcommand{\ETHZ}{Institute for Particle Physics, ETH Z\"urich, Z\"urich 8093, Switzerland}
\newcommand{\FNALaddress}{Fermi National Accelerator Laboratory, Batavia, IL 60510, USA}
\newcommand{\FortLewis}{Department of Physics and Engineering, Fort Lewis College, Durango, CO 81301, USA}
\newcommand{\GEUni}{Physics Department, Universit\`a degli Studi di Genova, Genova 16146, Italy}
\newcommand{\GEINFN}{INFN Genova, Genova 16146, Italy}
\newcommand{\Hawaii}{Department of Physics and Astronomy, University of Hawai'i, Honolulu, HI 96822, USA}
\newcommand{\Houston}{Department of Physics, University of Houston, Houston, TX 77204, USA}
\newcommand{\IHEPaddress}{Institute of High Energy Physics, Beijing 100049, China}
\newcommand{\IPNO}{Institut de Physique Nucl\`eaire d'Orsay, 91406, Orsay, France}
\newcommand{\INSTM}{Interuniversity Consortium for Science and Technology of Materials, Firenze 50121, Italy}
\newcommand{\IPHC}{IPHC, Universit\'e de Strasbourg, CNRS/IN2P3, Strasbourg 67037, France}
\newcommand{\JINR}{Joint Institute for Nuclear Research, Dubna 141980, Russia}
\newcommand{\Krakow}{M.~Smoluchowski Institute of Physics, Jagiellonian University, 30-348 Krakow, Poland}
\newcommand{\Kurchatov}{National Research Centre Kurchatov Institute, Moscow 123182, Russia}
\newcommand{\Laurentian}{Department of Physics and Astronomy, Laurentian University, Sudbury, ON P3E 2C6, Canada}
\newcommand{\Lancaster}{Physics Department, Lancaster University, Lancaster LA1 4YB, UK}
\newcommand{\Liverpool}{Department of Physics, University of Liverpool, The Oliver Lodge Laboratory, Liverpool L69 7ZE, UK}
\newcommand{\LNFINFN}{INFN Laboratori Nazionali di Frascati, Frascati 00044, Italy}
\newcommand{\LNLINFN}{INFN Laboratori Nazionali di Legnaro, Legnaro (Padova) 35020, Italy}
\newcommand{\Lodz}{Institute of Applied Radiation Chemistry, Lodz University of Technology, 93-590 Lodz, Poland}
\newcommand{\LPNHE}{LPNHE, CNRS/IN2P3, Sorbonne Universit\'e, Universit\'e Paris Diderot, Paris 75252, France}
\newcommand{\Mainz}{Institut f\"ur Kernphysik, Johannes Gutenberg-Universit\"at Mainz, D-55128 Mainz, Germany}
\newcommand{\Manchester}{Department of Physics and Astronomy, The University of Manchester, Manchester M13 9PL, UK}
\newcommand{\MEPhI}{National Research Nuclear University MEPhI, Moscow 115409, Russia}
\newcommand{\MendeleevUniverisity}{Mendeleev University of Chemical Technology, Moscow 125047, Russia}
\newcommand{\MIBIINFN}{INFN Milano Bicocca, Milano 20126, Italy}
\newcommand{\MIINFN}{INFN Milano, Milano 20133, Italy}
\newcommand{\MIPoliICA}{Civil and Environmental Engineering Department, Politecnico di Milano, Milano 20133, Italy}
\newcommand{\MIPoliCHE}{Chemistry, Materials and Chemical Engineering Department ``G.~Natta", Politecnico di Milano, Milano 20133, Italy}
\newcommand{\MIPoliEIB}{Electronics, Information, and Bioengineering Department, Politecnico di Milano, Milano 20133, Italy}
\newcommand{\MIPoliENE}{Energy Department, Politecnico di Milano, Milano 20133, Italy}
\newcommand{\MIUni}{Physics Department, Universit\`a degli Studi di Milano, Milano 20133, Italy}
\newcommand{\MSU}{Skobeltsyn Institute of Nuclear Physics, Lomonosov Moscow State University, Moscow 119234, Russia}
\newcommand{\NAINFN}{INFN Napoli, Napoli 80126, Italy}
\newcommand{\NAUniPHY}{Physics Department, Universit\`a degli Studi ``Federico II'' di Napoli, Napoli 80126, Italy}
\newcommand{\NAUniCHE}{Chemical, Materials, and Industrial Production Engineering Department, Universit\`a degli Studi ``Federico II'' di Napoli, Napoli 80126, Italy}
\newcommand{\NAUniPHARM}{Pharmacy Department, Universit\`a degli Studi ``Federico II'' di Napoli, Napoli 80131, Italy}
\newcommand{\NAUniStruct}{Department of Strutture per l'Ingegneria e l'Architettura, Universit\`a degli Studi ``Federico II'' di Napoli, Napoli 80131, Italy}
\newcommand{\NAUniEEIT}{Department of Electrical Engineering and Information Technology, Universit\`a degli Studi ``Federico II'' di Napoli, Napoli 80125, Italy}
\newcommand{\NSU}{Novosibirsk State University, Novosibirsk 630090, Russia}
\newcommand{\OACINAF}{INAF Osservatorio Astronomico di Capodimonte, 80131 Napoli, Italy}
\newcommand{\Petersburg}{Saint Petersburg Nuclear Physics Institute, Gatchina 188350, Russia}
\newcommand{\PGUniCBB}{Chemistry, Biology and Biotechnology Department, Universit\`a degli Studi di Perugia, Perugia 06123, Italy}
\newcommand{\PGINFN}{INFN Perugia, Perugia 06123, Italy}
\newcommand{\PIINFN}{INFN Pisa, Pisa 56127, Italy}
\newcommand{\PIUniPHY}{Physics Department, Universit\`a degli Studi di Pisa, Pisa 56127, Italy}
\newcommand{\PNNLaddress}{Pacific Northwest National Laboratory, Richland, WA 99352, USA}
\newcommand{\Princeton}{Physics Department, Princeton University, Princeton, NJ 08544, USA}
\newcommand{\Queens}{Department of Physics, Engineering Physics and Astronomy, Queen's University, Kingston, ON K7L 3N6, Canada}
\newcommand{\RHUL}{Department of Physics, Royal Holloway University of London, Egham TW20 0EX, UK}
\newcommand{\RMTreINFN}{INFN Roma Tre, Roma 00146, Italy}
\newcommand{\RMTreUni}{Mathematics and Physics Department, Universit\`a degli Studi Roma Tre, Roma 00146, Italy}
\newcommand{\RMUnoINFN}{INFN Sezione di Roma, Roma 00185, Italy}
\newcommand{\RMUnoUni}{Physics Department, Sapienza Universit\`a di Roma, Roma 00185, Italy}
\newcommand{\SAINFN}{INFN Salerno, Salerno 84084, Italy}
\newcommand{\SAUni}{Physics Department, Universit\`a degli Studi di Salerno, Salerno 84084, Italy}
\newcommand{\SNOLABaddress}{SNOLAB, Lively, ON P3Y 1N2, Canada}
\newcommand{\SSUniCHP}{Chemistry and Pharmacy Department, Universit\`a degli Studi di Sassari, Sassari 07100, Italy}
\newcommand{\STFCInterconnect}{Science \& Technology Facilities Council (STFC), Rutherford Appleton Laboratory, Technology, Harwell Oxford, Didcot OX11 0QX, UK}
\newcommand{\Sussex}{Physics and Astronomy Department, University of Sussex, Brighton BN1 9QH, UK}
\newcommand{\Temple}{Physics Department, Temple University, Philadelphia, PA 19122, USA}
\newcommand{\TNFBK}{Fondazione Bruno Kessler, Povo 38123, Italy}
\newcommand{\TNTIFPA}{Trento Institute for Fundamental Physics and Applications, Povo 38123, Italy}
\newcommand{\TNUni}{Physics Department, Universit\`a degli Studi di Trento, Povo 38123, Italy}
\newcommand{\TOINFN}{INFN Torino, Torino 10125, Italy}
\newcommand{\TOPoli}{Department of Electronics and Communications, Politecnico di Torino, Torino 10129, Italy}
\newcommand{\TOUni}{Physics Department, Universit\`a degli Studi di Torino, Torino 10125, Italy}
\newcommand{\TRIUMFaddress}{TRIUMF, 4004 Wesbrook Mall, Vancouver, BC V6T 2A3, Canada}
\newcommand{\TUM}{Physik Department, Technische Universit\"at M\"unchen, Munich 80333, Germany}
\newcommand{\UB}{Universiatat de Barcelona, Barcelona E-08028, Catalonia, Spain} 
\newcommand{\UCDavis}{Department of Physics, University of California, Davis, CA 95616, USA}
\newcommand{\UCRiverside}{Department of Physics and Astronomy, University of California, Riverside, CA 92507, USA}
\newcommand{\UCSanDiego}{Department of Physics, University of California, San Diego, CA 92093, USA}
\newcommand{\UCLA}{Physics and Astronomy Department, University of California, Los Angeles, CA 90095, USA}
\newcommand{\UCAS}{University of Chinese Academy of Sciences, Beijing 100049, China}
\newcommand{\UMass}{Amherst Center for Fundamental Interactions and Physics Department, University of Massachusetts, Amherst, MA 01003, USA}
\newcommand{\UNAM}{Instituto de F\'isica, Universidad Nacional Aut\'onoma de M\'exico, M\'exico 01000, Mexico}
\newcommand{\UnivAQ}{Universit\`a degli Studi dell’Aquila, L’Aquila 67100, Italy}
\newcommand{\UniversityofEdinburgh}{School of Physics and Astronomy, University of Edinburgh, Edinburgh EH9 3FD, UK}
\newcommand{\UOC}{Department of Chemistry, University of Crete, P.O. Box 2208, 71003 Heraklion, Crete, Greece}
\newcommand{\USP}{Instituto de F\'isica, Universidade de S\~ao Paulo, S\~ao Paulo 05508-090, Brazil}
\newcommand{\VTech}{Virginia Tech, Blacksburg, VA 24061, USA}
\newcommand{\Warwick}{University of Warwick, Department of Physics, Coventry CV47AL, UK}
\newcommand{\WilliamsCollege}{Williams College, Physics Department, Williamstown, MA 01267 USA}
\newcommand{\Zaragoza}{Centro de Astropart\'iculas y F\'isica de Altas Energ\'ias, Universidad de Zaragoza, Zaragoza 50009, Spain}
\newcommand{\ZaragozaARAID}{Fundaci\'on ARAID, Universidad de Zaragoza, Zaragoza 50009, Spain}

\author{P.~Agnes}\affiliation{\RHUL}\affiliation{\TRIUMFaddress}
\author{I.~Ahmad}\affiliation{\AstroCeNT}
\author{S.~Albergo}\affiliation{\CTINFN}\affiliation{\CTUNI}
\author{I.~F.~M.~Albuquerque}\affiliation{\USP}
\author{T.~Alexander}\affiliation{\PNNLaddress}
\author{A.~K.~Alton}\affiliation{\Augustana}
\author{P.~Amaudruz}\affiliation{\TRIUMFaddress}
\author{M.~Atzori Corona}\affiliation{\CAUniPHY}\affiliation{\CAINFN}
\author{D.~J.~Auty}\affiliation{\Alberta}
\author{M.~Ave}\affiliation{\USP}
% ...
\collaboration{Global Argon Dark Matter Collaboration}\email{ds-ed@lngs.infn.it}\noaffiliation

\date{\today}
\begin{abstract}
Dark matter lighter than \SI{10}{\GeV\per\square\c} encompasses a promising range of candidates.
A conceptual design for a new detector, \DSl, is presented, based on the \DSf\ detector and progress toward \DSk, optimized for a low-threshold electron-counting measurement.
Sensitivity to light dark matter is explored for various potential energy thresholds and background rates.
These studies show that \DSl\ can achieve sensitivity to light dark matter down to the solar neutrino fog for \si{\GeV}-scale masses and significant sensitivity down to \SI{10}{\MeV\per\square\c} considering the Migdal effect or interactions with electrons. 
Requirements for optimizing the detector's sensitivity are explored, as are potential sensitivity gains from modeling and mitigating spurious electron backgrounds that may dominate the signal at the lowest energies.
\end{abstract}
\maketitle

\section{Introduction}
Astrophysical evidence indicates that dark matter (\DM) constitutes \SI{26}{\percent} of the universe's energy density~\cite{planckcollaborationPlanck2018Results2020}.
Many experiments have tried to detect it directly, often focused on Weakly Interacting Massive Particles (\WIMPs) with mass between \WIMPMassLowLimit\ and \WIMPMassTenTev~\cite{darkside_collaboration_results_2016,deap_collaboration_constraints_2020, aprile_dark_2018, lux_collaboration_results_2017, pandax-ii_collaboration_dark_2017,supercdms_collaboration_results_2018,pico_collaboration_dark_2017}.
Planned experiments~\cite{akeribSnowmass2021CosmicFrontier2022,aalbersDARWINUltimateDark2016,aalsethDarkSide20k20Tonne2018} will search for \WIMPs\ with cross sections below which Coherent Elastic Neutrino-Nucleus Scattering (\CEnNS) from atmospheric neutrinos may obscure \DM\ signals, called the ``neutrino fog''~\cite{billard_implication_2014,ohareNewDefinitionNeutrino2021}.

Past experiments show that similar technology can perform dedicated light \DM\ searches~\cite{the_darkside_collaboration_constraints_2018,darkside_collaboration_low-mass_2018,aprile_light_2019,akerib_results_2018,alkhatib_light_2021,mengDarkMatterSearch2021}.
\DSf\ demonstrated that a dual-phase liquid argon time-projection chamber (\LArTPC) performing an electron-counting analysis---focused on electroluminescence signals from ionization electrons in a gas pocket, \STwo---is sensitive to \DM\ with nuclear couplings for \SIrange{1}{10}{\GeV\per\square\c} masses~\cite{ds50SearchLowmassDark2022,ds50Migdal2022} and electronic couplings for \SIrange{0.01}{1}{\GeV\per\square\c} masses~\cite{ds50electronic2022}.

Interactions in \LAr\ produce a comparable amount of scintillation and ionization. 
While photons are detected with \SI{\sim20}{\percent} efficiency and must overcome noise, the near-perfect efficiency for extracting electrons from liquid to gaseous argon~\cite{bondarElectronEmissionProperties2009}, the long drift lifetime (enabled by excellent purity achievable in \LAr), and gas pocket amplification lets each ionization electron be detected.
As a result, the electron-counting channel accesses energies near the work function, lower than those reached by scintillation.

Dual-phase \LArTPCs\ benefit from scalability due to \LAr's high transparency to photons and electrons; their low temperature enables exceptional purity, as seen in \DEAP's low \isotope{222}{Rn} concentration~\cite{deap_collaboration_search_2019} and \DSf's long electron drift lifetime~\cite{darkside_collaboration_darkside-50_2018}.
The relatively light nucleus also allows light \DM\ to produce higher-energy recoils.

These properties enable dual-phase \LArTPCs\ to search for light \DM\ down to the neutrino fog with index n$>$1.5. 
Maximizing sensitivity requires a dedicated detector optimized for electron-counting analyses by enhancing \STwo\ and minimizing backgrounds that produce \SI{<3}{\keV} electron equivalent (\si{\keVee}) signals, as expected from light \DM.
\DSl\ aims to employ such a detector.
This paper explores its potential sensitivity, considering \num{2} and \SI{4}{\el} analysis thresholds and possible background levels and detector response models.
A conceptual design is presented in \refsec{sec:design}; \refsec{sec:model} describes response models, and \refsec{sec:background} explores background scenarios. 
Finally, \refsec{sec:sensitivity} projects sensitivity with these models, and \refsec{sec:improvements} discusses potential future improvements.

\section{Conceptual detector design}\label{sec:design}
\begin{figure}[htb]
    \centering
    \includegraphics[width=\linewidth]{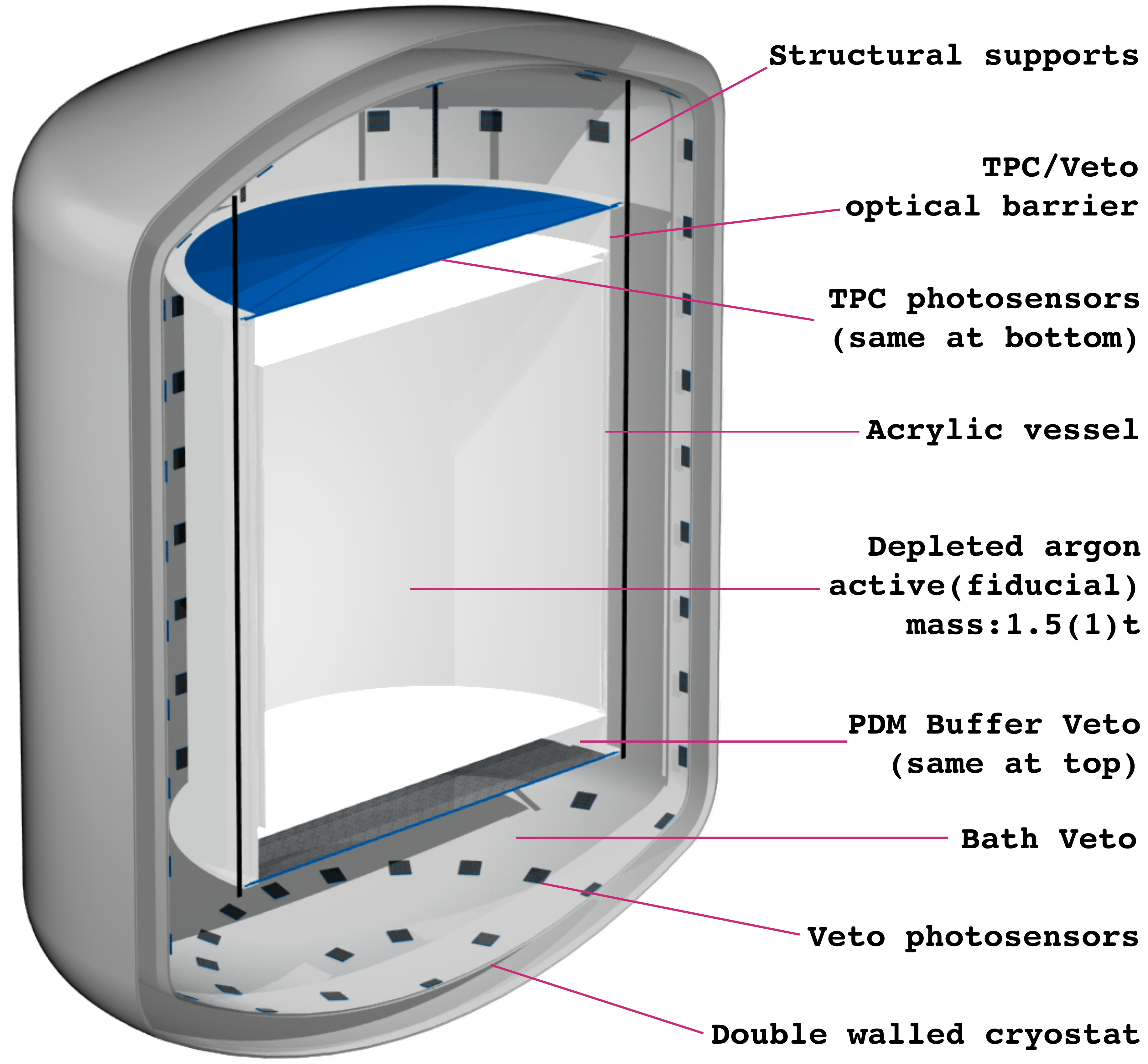}
    \caption{Conceptual detector design: a \DSLMTargetActiveMass\ dual-phase \LArTPC\ in an acrylic vessel, viewed by two photosensor arrays via \DSLMPDMOffset\ ``buffer vetoes'', in a \UAr\ ``bath veto'' in a cryostat, immersed in a water tank (not shown). 
    }
    \label{fig:detector_dwg}
\end{figure}

Based on lessons from \DSf\ and progress toward \DSk~\cite{darkside_collaboration_darkside-50_2018, darkside_collaboration_low-mass_2018, the_darkside_collaboration_constraints_2018, ds50SearchLowmassDark2022,ds50Migdal2022,ds50electronic2022, darkside_collaboration_separating_2021,the_darkside_collaboration_calibration_2021,collaboration_study_2021,aalsethCryogenicCharacterizationFBK2017}, a conceptual detector has been designed to optimize \DSl\ for low-threshold analyses.

\subsection{Lessons from \DSf}
\label{subsec:lessons}

While \DSf\ was designed for a high-mass \WIMP\ search using primary scintillation (\SOne) and electroluminescence (\STwo), its sensitivity to light \DM\ elucidates how a dual-phase \LArTPC\ can be optimized for an electron-counting analysis.
This channel lacks \SOne, thereby losing the capacity to reject electronic recoils (\ERs) by pulse shape discrimination and to reconstruct interactions' vertical positions~\cite{essigSnowmass2021CosmicFrontier2022}.
\DSf's sensitivity was limited by \ERs\ due to \grs\ from the photomultiplier tubes and cryostat and \bta-decays of trace residual \isotope{85}{Kr} and \isotope{39}{Ar} in the argon extracted from underground (\UAr)~\cite{backFirstLargeScale2012,backFacilityLowRadioactivityUnderground2022}.
At the lowest energies, spurious electrons (\SEs), not directly produced by energy depositions, dominate \SI{<4}{\el} backgrounds, imposing an effective analysis threshold.
Mitigating \SEs\ is key to improving \DSl's sensitivity.

\subsection{Detector description}
\label{subsec:design}

\begin{table}[htb]
    \centering
    \caption{Conceptual detector design parameters.}
    \begin{tabular}{l|r}\hline\hline 
        Parameter                   & Value                 \\\hline
        \TPC\ active \LAr\ mass     & \DSLMTargetActiveMass       \\
        \TPC\ fiducial \LAr\ mass   & \DSLMTargetFiducialMass     \\
        \TPC\ fiducial cylindrical radius& \DSLMTargetFiducialRadius   \\
        \TPC\ height                & \DSLMTPCHeight              \\
        \TPC\ diameter              & \DSLMTPCDiameter            \\
        \TPC\ \PDM\ number          & \DSLMNumPDMs                \\
        \TPC\ \PDM\ peak efficiency & \DSLMPeakPDEEfficiency \\
        \TPC\ gas pocket thickness  & \DSLMGasPocketThickness     \\ 
        \TPC\ electroluminescence field & \DSLMELField           \\
        \TPC\ drift field           & \DSLMDriftField             \\ 
        Acrylic vessel mass         & \SI{0.144}{\tonne} \\
        PDM dimensions              & \SI[product-units=single]{5x5}{\square\cm} \\    
        PDM buffer veto thickness   & \DSLMPDMOffset              \\
        PDM buffer veto total mass  & \SI{0.3}{\tonne}            \\
        Bath veto \UAr\ mass        & \DSLMBathVetoUArMass        \\ 
        Bath veto minimum thickness & \DSLMBathVetoThickness      \\
        Cryostat inner height       & \DSLMCryostatHeight         \\
        Cryostat inner diameter     & \DSLMCryostatDiameter       \\
        Cryostat wall thickness     & \DSLMCryostatWallThickness  \\
        Ti support structure total mass & \SI{0.1}{\tonne} \\
        \hline\hline
    \end{tabular}
    \label{tab:detector_parameters}
\end{table}

Figure~\ref{fig:detector_dwg} shows a conceptual \DSl\ design; \reftab{tab:detector_parameters} gives design parameters.
The nested structure isolates and vetoes against radioactivity.
The detector consists of the following elements:

\paragraph{Depleted Argon TPC:}
the inner detector is a dual-phase \TPC\ with an active (fiducial) mass of \SI{1.5}{\tonne} (\SI{1}{\tonne}) of \UAr, depleted of \isotope{39}{Ar} by cryogenic distillation~\cite{darkside_collaboration_separating_2021}. 
The \TPC\ has an ultra-pure acrylic vessel, as in \DEAP~\cite{amaudruz_design_2019}.
Transparent conductive films like \Clevios\ define anode and cathode planes; rings coated on the walls ensure spatial uniformity of the drift field. 
Electroluminescence in a \DSLMGasPocketThickness-thick gas pocket at the top allows extracted electrons to be counted.
A stainless steel grid below the \LAr\ surface separates the drift volume from the extraction and multiplication regions, with a \DSLMDriftField\ drift field in the bulk and a \DSLMELField\ electroluminescence field in the gas pocket, building on experience from \DSf.
The expected extraction efficiency exceeds \SI{99.9}{\percent}~\cite{bondarElectronEmissionProperties2009}.

The vessel's inner surfaces are lined with reflector coated with wavelength shifter like \TPB\ (tetraphenyl butadiene), which shifts VUV photons emitted by argon to \SI{\sim420}{\nm}. 
Two planes of photodetector modules (\PDMs) with $100\%$ optical coverage, mounted \DSLMPDMOffset\ above and below the \TPC, detect this light.
Each \PDM\ is a \SI[product-units=single]{5x5}{\square\cm} array of silicon photomultipliers (\SiPMs) based on \refcite{dinceccoDevelopmentNovelSingleChannel2018}, readout by cryogenic pre-amplifiers developed for \DSk~\cite{dinceccoDevelopmentVeryLownoise2018}.
Titanium structural supports hold the \TPC\ and optical planes; 
titanium allows them to be radiopure and lightweight, reducing their impact on the vetoes and background budget.
This system is immersed in a \UAr\ bath held in a double-walled, \DSLMCryostatDiameter-diameter stainless steel cryostat.

For these studies, the \TPC\ has equal diameter and height in order to maximize the path that external \grs\ must traverse before reaching the fiducial volume.
This design also balances the inability to fiducialize along the vertical axis against the longer electron drift time in a taller \TPC, which requires longer veto windows and higher voltages.
Considering possible effects of drift time on \SEs\ (see \refsec{subsec:ses}), other designs may be motivated by future work.

\paragraph{\gr\ vetoes:} the \TPC\ is surrounded by two vetoes. 
These instrumented \LAr\ volumes provide passive buffers and  anti-coincidence signals when \grs\ deposit energy in them before or after scattering in the \TPC.
\textbf{\PDM\ buffer veto:} Reflective and wavelength-shifting foils (\eg\ TPB-coated ESR and acrylic surfaces) surround both \PDM\ arrays and the acrylic vessel, optically decoupling them from the \LAr\ bath while enhancing light collection efficiency.
The \SI{10}{\cm} offset between each optical plane and the acrylic vessel serves as a ``buffer'' veto for \grs\ emitted by the \PDMs\ and associated hardware. 
This offset allows \si{\cm}-scale spatial resolution, and larger offsets marginally impact the background rate.
Veto scintillation is separated from \SOne\ and \STwo\ in the \TPC\ by pulse-shape and the concentration of light in either \PDM\ plane.
\textbf{Bath veto:} The \DSLMBathVetoUArMass\ \UAr\ in the cryostat is instrumented with \PDMs\ on the cryostat walls and functions as another \gr\ veto. 
This \DSLMBathVetoThickness\ buffer uses the minimal \UAr\ mass needed to veto and shield against \grs\ from the cryostat and render their backgrounds subdominant.

\paragraph{Water shielding:}
the cryostat is in a \SIrange{8}{10}{\meter}-diameter water tank that shields against external radiation. 
If a cosmic-ray veto is needed, the tank can be instrumented to detect Cherenkov light.

\section{Detector response model}\label{sec:model}

The detector response model closely follows that in \refscite{bezrukov_interplay_2011,the_darkside_collaboration_calibration_2021}. For a nuclear recoil (\NR) of energy $E_R$, this model uses the reduced energy $\epsilon$, defined as
\begin{equation}
    \epsilon=0.626\frac{a_0}{e^2}\frac{E_R}{2Z^{7/3}}
\end{equation}
with target atomic number $Z$, Bohr radius $a_0$, and elementary charge $e$, giving \mbox{$a_0/e^2=\SI{36.81}{\per\keV}$}.

\begin{figure}[ht]
    \centering
    \includegraphics[width=\linewidth]{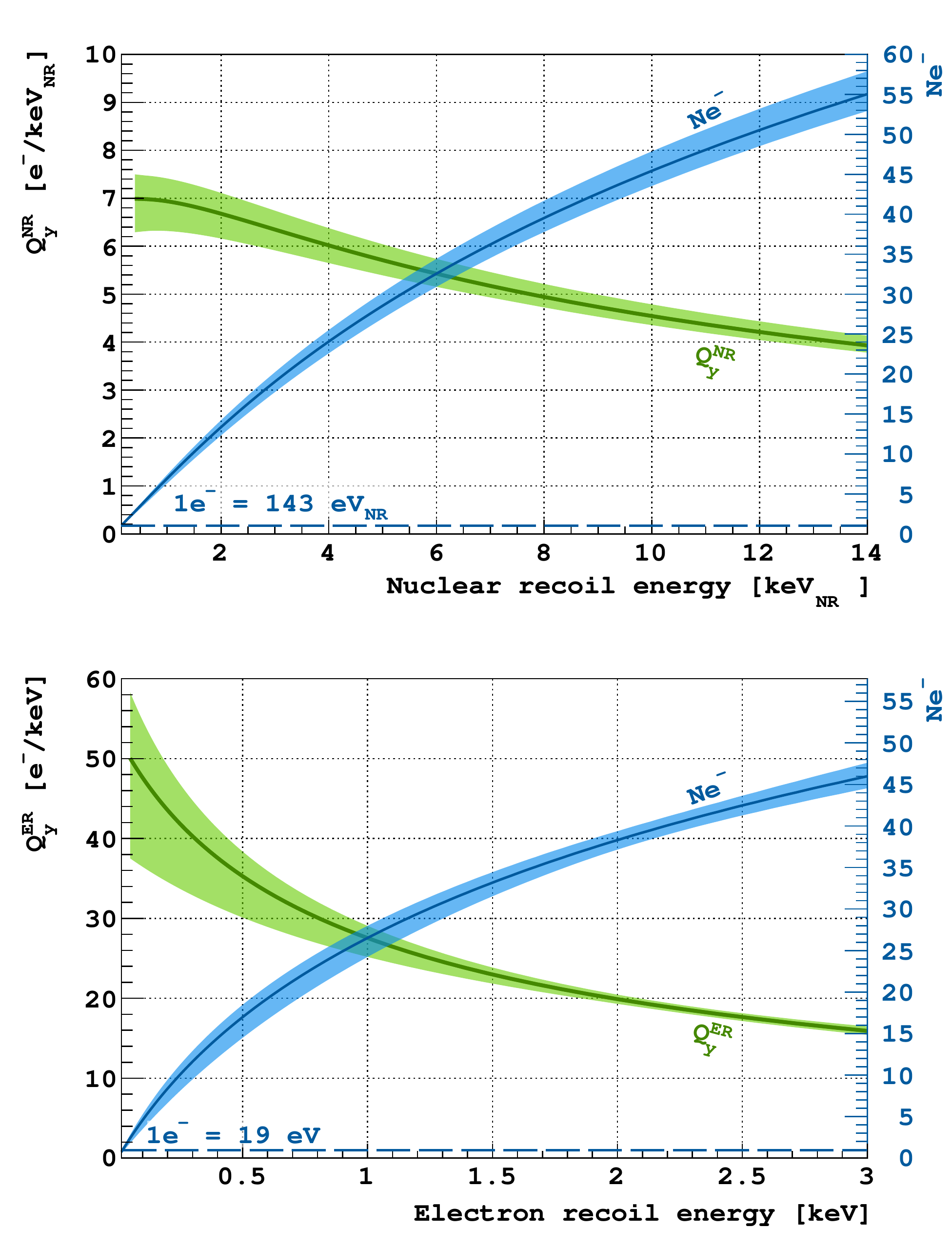}
    \caption{Ionization yield models assumed in these studies for (top) nuclear recoils and (bottom) electronic recoils. Bands show $\pm1\sigma$ uncertainty from the fit to constraints.}
    \label{fig:qy}
\end{figure}

The number of ionization electrons that escape recombination and contribute to \STwo\ is given by
\begin{equation}
    \NeNR = \frac{\DriftFieldSymbolPow}{\cBoxNR} \ln\left|1+\frac{10^4\fB\cBoxNR}{\DriftFieldSymbolPow} \frac{\epsilon \StoppingPowerElec(\epsilon)}{\StoppingPowerElec(\epsilon)+\StoppingPowerNucl(\epsilon)}\right| ;
\end{equation}
\fB\ describes the energy partition among ionization and other modes, \cBoxNR\ is a parameter describing the spatial extent of the \NR\ charge distribution, $\DriftFieldSymbol$ is the drift field strength in \si{V/cm}, $B$ parameterizes the drift field scaling, and \StoppingPowerElec\ and \StoppingPowerNucl\ are the electronic and nuclear stopping powers, given by Zielger as~\cite{ziegler_stopping_1985}
\begin{gather}
    \begin{aligned}
    S_e &= 0.133\frac{Z^{2/3}}{A^{1/2}}\sqrt{\epsilon} \\
    S_n &= \frac{\ln{\left|1. + 1.1383\epsilon_Z\right|}}{2\left(\epsilon_Z+0.01321\epsilon_Z^{0.21226}+0.19593\epsilon_Z^{0.5}\right)}
    \end{aligned}
\end{gather}
where $\epsilon_Z=0.94\epsilon$ for argon, accounting for atomic screening effects, and $A$ is the mass number of argon. 
\refcite{the_darkside_collaboration_calibration_2021} found that this model consistently gives lower \ChargeYieldNR\ than those by Moli\`ere~\cite{moliereTheorieStreuungSchneller1947} and by Lenz and Jensen~\cite{lenzUeberAnwendbarkeitStatistischen1932,jensenLadungsverteilungIonenUnd1932}, making its use conservative.

For \ERs, the number of electrons escaping recombination is described by the Thomas-Imel model as
\begin{equation}
    \NeER =  \frac{\DriftFieldSymbolPow}{\cBoxER}\ln{\left|1+\frac{\cBoxER}{\DriftFieldSymbolPow}\rho E_R\right|} + 1
\end{equation}
where the initial recoiling electron has been added to $N_e^\text{ER}$, and $\rho$, \cBoxER, and $B$ are model parameters. 

\begin{table}[htb]
    \centering
    \renewcommand{\arraystretch}{1.2}
    \caption{(Top) \ChargeYieldNR\ and \ChargeYieldER\ and (Bottom) optical model parameters. Fixed parameters are shown with their assumed values; fit parameters are shown with their best-fit values to external data and the range over which they were allowed to float in sensitivity projections.}
    \begin{tabular}{c|ccc}\hline\hline 
    & \multicolumn{3}{c}{Charge yield parameters}\\
         & Bounds & Modeled value   & Units                  \\\hline 
        \DriftFieldSymbol & Fixed   & \num{200}   & \si{\volt\per\cm} \\
        \EFieldPower      & Fixed   & \num{0.61}  & ---               \\
        \cBoxNR           & \SIrange[range-phrase={,}\ ,range-units= brackets,open-bracket=[,close-bracket=]]{0.51}{2.04}{\,} &  \num{1.02}$_{-0.03}^{+0.01}$ & $(\si{\volt\per\cm})^B$ \\
        \fB               & 
        \SIrange[range-phrase={,}\ ,range-units= brackets,open-bracket=[,close-bracket=]]{0.35}{1.38}{\,} & \num{0.69}$_{-0.05}^{+0.04}$ & --- \\ 
        \cBoxER           & \SIrange[range-phrase={,}\ ,range-units= brackets,open-bracket=[,close-bracket=]]{0.55}{2.18}{\,} &  \num{1.09}$_{-0.20}^{+0.19}$ & $(\si{\volt\per\cm})^B$ \\
        $\rho$            & \SIrange[range-phrase={,}\ ,range-units= brackets,open-bracket=[,close-bracket=]]{27}{106}{\,} & \num{53}$_{-10}^{+12}$ & \si[per-mode=reciprocal]{\per\keV}  \\\hline\hline 
        & \multicolumn{3}{c}{\STwo\ response parameters} \\
         & Modeled value & Units & \\\hline 
        \CollectionEfficiencySymbol &  \DSLMCollectionEfficiencyNum      & \si{\pe\per\ph} & \\
        \STwoPhotonYieldSymbol     & \DSLMSTwoPhotonYieldNum         & \si{\ph\per\el} & \\
        \gTwo                & \DSLMgTwoNominalNum          & \si{\pe\per\el} & \\ 
        \XYResolutionSymbol  &  \DSLMXYResolutionNum & \si{\cm} & \\
        \vdrift & \DSfLArBelowMeshElectronSpeedNum & \si{\mm\per\micro\second} \\
        \hline\hline 
    \end{tabular}
    \label{tab:model_params}
\end{table}

The charge yield is defined as $\ChargeYieldNR = \NeNR/E_R$ and $\ChargeYieldER = (\NeER-1)/E_R$. 
Nuisance parameters \fB, \cBoxER, \cBoxNR, and $\rho$ are constrained by \ChargeYieldNR\ and \ChargeYieldER\ measurements reported by \ARIS~\cite{ARIS} and \SCENE~\cite{scene_collaboration_measurement_2015}, as well as \insitu\ measurements reported by \DSf~\cite{darkside_collaboration_low-mass_2018,the_darkside_collaboration_calibration_2021}.
This treatment follows that in \refcite{the_darkside_collaboration_calibration_2021}, constrained to energies below \SI{3}{\keVee}.
Since the Thomas-Imel model is valid for \ERs\ in this full range, the extended model developed in \refcite{the_darkside_collaboration_calibration_2021} is not needed here.
Furthermore, $B$ is fixed to the central value reported by \SCENE\ of $B=0.61$.
Studies varying \DriftFieldSymbol\ between \DSLMDriftFieldStudyRange\ show that its effects on \ChargeYieldNRER\ do not impact the projected sensitivity due to the low recombination rate at these energies.
As a result, \DriftFieldSymbol\ can be optimized based on its influence on other parameters like \vdrift; such studies are left to future work.
\reffig{fig:qy} shows the models that best fit these constraints;  \reftab{tab:model_params} gives fixed and fit parameters.

Non-uniformities in \gTwo, as might arise from \TPC\ components sagging, will lower \SI{1}{\el} resolution.
These effects can be corrected using the event position, with adequate resolution achieved with \DSl's stronger electroluminescence field. 
These defects may also be minimized by adopting technology from $\mathcal{O}(\SI{10}{\tonne})$ detectors, which have wire grids and anode planes several times larger than \DSl\ will have.

Quenching and recombination \NeNR\ fluctuations are modeled with a binomial function; uncertainties in this treatment are explored by projections with no quenching fluctuations. 
Binomial \NeER\ fluctuations are suppressed by a Fano-like factor $F$, as in \NEST~\cite{szydagisReviewBasicEnergy2021}, constrained by fits to \ce{^37Ar} peaks in \refcite{the_darkside_collaboration_calibration_2021}.
\refcite{the_darkside_collaboration_calibration_2021} assumes Gaussian fluctuations with variance $F\Ne$, valid for \Ne\SI{>10}{\el}. 
Lacking lower-energy \ER\ calibration, assumptions are needed to extrapolate below \SI{10}{\el}.
A binomial model reflects variations of energy dissipation via ionization and other modes, while a Gaussian model may describe variations in energy transferred to ionization electrons.
Both models fit \DSf's \ce{^37Ar} peaks.
Due to the strong \SI{1}{\el} resolution assumed for \DSl, the Gaussian model produces larger fluctuations at low \Ne, giving up to \SI{10}{\times} stronger constraints on \DM\ scattering cross sections; \DSf's resolution was dominated by spatial \gTwo\ variations, where \gTwo\ is the \STwo\ gain factor, so these models marginally impact its analysis.
Binomial fluctuations in \NeER\ suppressed by $F$ are conservatively assumed in the present studies. 
The number of detected photoelectrons (\pe) is drawn from a Poisson distribution with mean \mbox{$\gTwo\times\NeNRER$}. 

\subsection{Optical simulations}\label{ssec:model:optical}

Optical simulations were performed with \GFDS~\cite{agnes_simulation_2017}, based on \Geant\texttt{-10.0}~\cite{agostinelliGeant4SimulationToolkit2003}. 
\SiPMs\ are assumed to be similar to those in \refscite{aalsethCryogenicCharacterizationFBK2017,golaNUVSensitiveSiliconPhotomultiplier2019a,acerbiCryogenicCharacterizationFBK2017a}, with peak photon detection efficiency (PDE) of \DSLMPeakPDEEfficiency, including an \SI{\sim88}{\percent} geometrical fill factor.
Lower PDE and optical coverage can be compensated with a stronger electroluminescence field, increasing \STwoPhotonYieldSymbol. 

Simulations of VUV photons generated uniformly in the gas pocket predict a \STwo\ light collection efficiency of \mbox{$\CollectionEfficiencySymbol=\DSLMCollectionEfficiency$} with both \PDM\ arrays.
Electroluminescence simulations based on \refscite{zhu_electroluminescence_2018,oliveiraSimulationToolkitElectroluminescence2011} predict a photon yield of \mbox{$\STwoPhotonYieldSymbol=\DSLMSTwoPhotonYield$}.
These values give  \mbox{$\gTwo=\CollectionEfficiencySymbol\times\STwoPhotonYieldSymbol=\DSLMgTwoNominal$}, tuned for high \SI{1}{\el} detection efficiency and resolution, based on \DSf.
By varying the gas pocket thickness and electroluminescence field, \gTwo\ can be varied as needed.
These values are summarized in \reftab{tab:model_params}.

To estimate the horizontal position resolution,
\SI{\leq0.3}{\keV} electrons are simulated uniformly in the \LAr.
\STwo\ photons are generated in the gas pocket for each extracted electron, offset by \tdrift, the time required to drift electrons from the interaction vertex to the gas pocket.
The \STwo\ pulse shape is described in \refcite{agnes_electroluminescence_2018}. 
Photons that reach the \PDMs\ are registered as \pe\ with probability governed by the \PDE. 

\begin{figure}[htb]
    \centering
    \includegraphics[width=\linewidth]{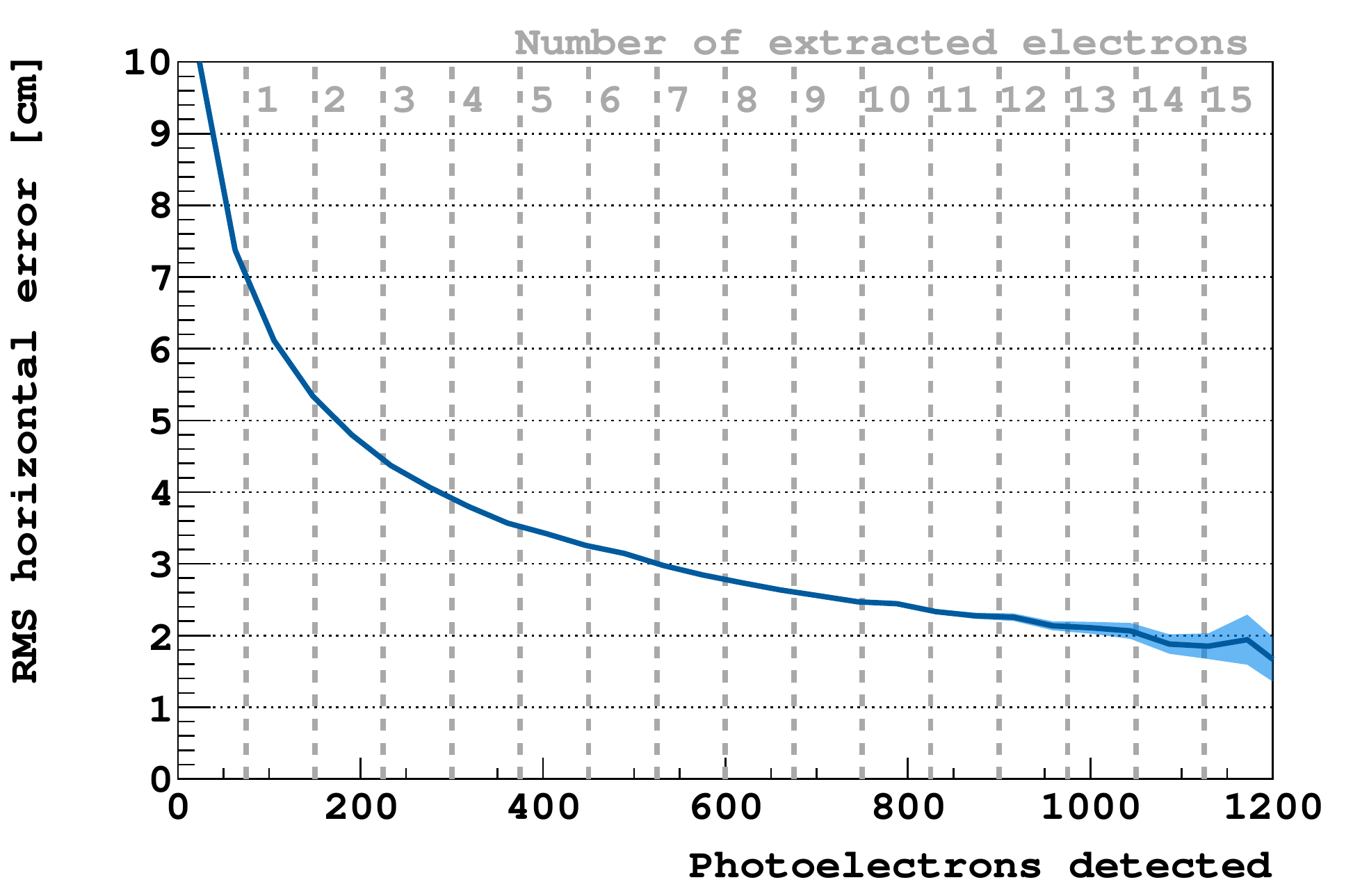}
    \caption{Horizontal position resolution \XYResolutionSymbol\ as a function of signal size. Between \SIrange{1}{15}{\el}, \XYResolutionSymbol\ is between \SIrange{2}{7}{\cm}. The dark line shows the median \XYResolutionSymbol, and the lighter band is the $1\sigma$ confidence belt from simulation statistics.}
    \label{fig:XY_resolution_0.3_0.7_1.1keV}
\end{figure}

The position of a signal in the horizontal plane is estimated using the barycenter method, calculated as the \pe-weighted average \PDM\ location and corrected for the expected radial bias near the walls.
More sophisticated algorithms can achieve better resolution, as demonstrated in \DSf~\cite{brodsky_xy_2015}.

The resolution \XYResolutionSymbol\ is defined as the root mean square (RMS) distance between the reconstructed and true positions. 
\reffig{fig:XY_resolution_0.3_0.7_1.1keV} shows \XYResolutionSymbol\ as a function of \STwo\ charge.
Overall, \XYResolutionSymbol\ decreases for larger signals, varying from \SIrange{2}{7}{\cm} for \SI{\geq1}{\el} signals.
For background simulations, a nominal resolution of $\XYResolutionSymbol=\DSLMXYResolution$ is assumed, though varying its value within this range does not impact the results presented.
While more sophisticated algorithms can likely achieve better resolution, varying the size of the \PDMs\ does not have a significant effect.

\section{Background model predictions}\label{sec:background}

At the highest energies relevant for light \DM, the primary backgrounds include $\gamma$-rays from detector components, $\beta$-decays from \LAr\ radioisotopes, cosmogenic activation of detector materials, surface backgrounds, and neutrinos. 
At the lowest energies, \SEs\ produce the dominant backgrounds.

\subsection{Neutrinos}
\begin{figure}[htb]
    \centering
    \includegraphics[width=\linewidth]{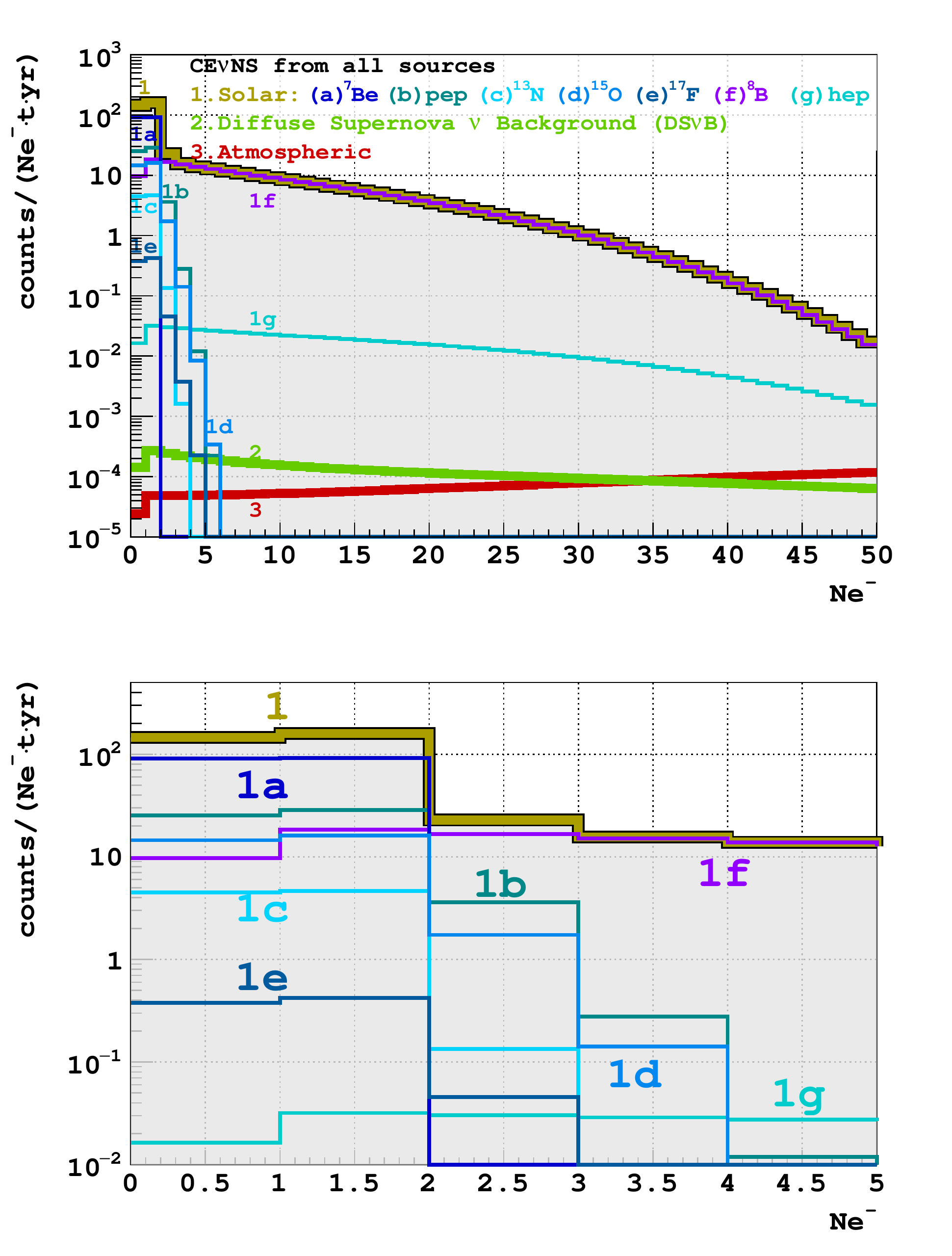}
    \caption{Coherent elastic neutrino-nucleus scattering background from all sources ($pp$ neutrinos not visible), zoomed into \emph{(top)} \SIrange{0}{50}{\el} and \emph{(bottom)} \SIrange{0}{5}{\el}.}
    \label{fig:neutrinos}
\end{figure}
\CEnNS\ from solar and atmospheric neutrinos and the diffuse supernova neutrino background (DS$\nu$B) pose a currently-irreducible background.
Their fluxes are modeled as in \refcite{baxter_recommended_2021}, summarized in Table~\ref{tab:neutrino_flux}, giving \SI{498\pm12}{\ev\per\tonne\per\year} between \SIrange{1}{50}{\el} (about \SIrange{0.14}{12}{\keVr} or \SIrange{0.02}{3.4}{\keVee}), dominated by \isotope{8}{B} solar neutrinos above \SI{3}{\el} and by \ce{^7Be}, $pep$, and CNO neutrinos below.
Neutrino-electron scattering, mostly from $pp$ neutrinos, will produce \DSLMnuESBkgd\ in the same \Ne\ range.

\begin{table}[htb]
    \centering
    \caption{Neutrino fluxes assumed in these studies and their associated uncertainties. For solar neutrinos, the high metallicity model was assumed.}
    \begin{tabular}{l|ccc}\hline\hline 
           & Flux [\si{\per\square\cm\per\second}] & Uncertainty    & Ref.\\\hline 
        $pp$            & \num{5.98E10} & \SI{0.6}{\percent} & \cite{vinyoles_new_2017} \\
        $pep$           & \num{1.44E8}  & \SI{1}{\percent}   & \cite{vinyoles_new_2017}\\
        \isotope{7}{Be} & \num{4.99E9}  & \SI{3}{\percent}   & \cite{borexino_collaboration_simultaneous_2019}\\
        \isotope{8}{B}  & \num{5.25E6}  & \SI{4}{\percent}   & \cite{sno_collaboration_combined_2013}\\ 
        $hep$           & \num{7.98E3}  & \SI{30}{\percent}  & \cite{vinyoles_new_2017}\\ 
        \isotope{13}{N} & \num{2.78E8}  & \SI{15}{\percent}  & \cite{vinyoles_new_2017}\\
        \isotope{15}{O} & \num{2.05E8}  & \SI{17}{\percent}  & \cite{vinyoles_new_2017}\\ 
        \isotope{17}{F} & \num{5.29E6}  & \SI{20}{\percent}  & \cite{vinyoles_new_2017}\\
        Atmospheric     & \num{10.5}    & \SI{20}{\percent}  & \cite{battistoni_atmospheric_2005}\\
        DS$\nu$B            & \num{86}      & \SI{50}{\percent}  & \cite{keil_monte_2003}\\
        \hline\hline
    \end{tabular}
    \label{tab:neutrino_flux}
\end{table}

Figure~\ref{fig:neutrinos} shows the \CEnNS\ \NR\ spectra.
These irreducible backgrounds lead to the ``neutrino fog'': the \DM-nucleon cross section below which \CEnNS\ backgrounds impede sensitivity~\cite{billard_implication_2014,ohareNewDefinitionNeutrino2021}.
While solar neutrinos limit the \DM\ search, they enable solar neutrino studies.
The fog in Figs.~\ref{fig:sensitivity_scenarios} and~\ref{fig:nscat_discovery} is given for spectral indices $n=-(d\log\sigma_\text{SI}/d\log MT)^{-1}$, defining the gradient of the median spin-independent cross section $\sigma_\text{SI}$ that an experiment can observe at $3\sigma$ significance with exposure $MT$~\cite{ohareNewDefinitionNeutrino2021}.

\subsection{$\gamma$-ray backgrounds}

\begin{figure}[htb]
    \centering
    \includegraphics[width=\linewidth]{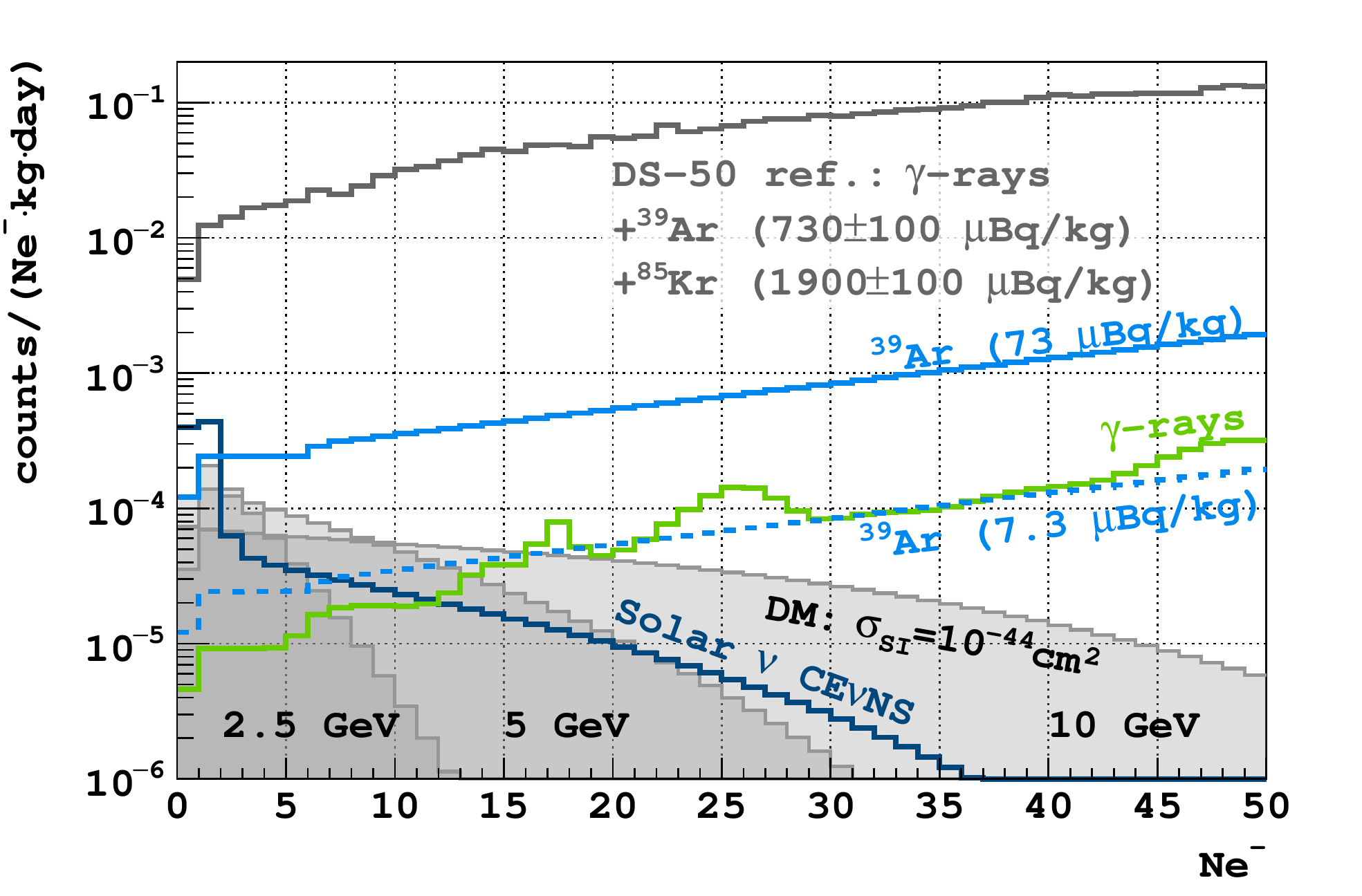}
    \caption{Backgrounds from \grs, \isotope{39}{Ar}, and \CEnNS, compared to \DSf. \DM\ spectra for are shown at \numlist{2.5;5;10}\,\si{\GeV\per\square\c} masses with spin-independent nucleon-scattering cross section \mbox{$\sigma_\text{SI}=\SI{e-44}{\square\cm}$}.}
    \label{fig:gammbgs}
\end{figure}

Radioisotopes emit \grs\ that scatter in the \TPC. 
Assays from \DEAP~\cite{amaudruz_design_2019} and \DS\ are used to estimate the activity of all detector components.
Dominant backgrounds include \xrs\ from the acrylic and \grs\ from the \PDMs, including photosensors and their hardware---mostly from \isotope{40}{K}
and the \ce{^238U} chain (\ce{^238U} to \ce{^230Th}).

Radioactive decays in all detector components were simulated using \GFDS~\cite{agnes_simulation_2017}.
Energy depositions were recorded in the \TPC, bath veto, and \PDM\ buffer veto, and the expected signals were reconstructed using the response model.
The electron drift time in the \TPC\ was determined by the drift speed and diffusion in \refcite{agnes_electroluminescence_2018}.
Events were rejected by a multiple-scatter cut if at least two \STwo\ signals were separated by \SI{>4}{\micro\second}.
The reconstructed position for events in the horizontal plane was determined using the barycenter coordinates smeared by a Gaussian to account for resolution. 
Varying the smearing within the range in \reffig{fig:XY_resolution_0.3_0.7_1.1keV} changes the observed background rate by \SI{<10}{\percent}.
Events outside of the inner \SI{1}{\tonne} core of the \TPC, defined in the horizontal plane, were rejected by a fiducial cut. 

Events are rejected if more than \DSLMBathVetoThreshold\ (\DSLMPDMBufferVetoThreshold) of energy is deposited in the bath (\PDM\ buffer) veto, within an anti-coincidence window of $\tdriftmax=\DSLMVetoWindow$ preceding the \STwo\ time. 
The use of UAr in the vetoes allows thresholds below the \isotope{39}{Ar} endpoint; accounting for energy depositions in both vetoes from \grs\ and \isotope{39}{Ar} decays, a \SIrange{3.5}{4.0}{\percent} dead time is expected, depending on \vdrift\ and the \isotope{39}{Ar} activity.

Total background rates after selection cuts are shown in \reffig{fig:gammbgs}, compared to \DSf's best-fit backgrounds and example \DM\ signals.
Following veto cuts, the total \gr\ background rate at $\Ne\SI{<12}{\el}$ is below that from solar neutrinos. 
No further R\&D is needed to improve \PDM\ radiopurity.

\subsubsection{Effects of the PDM buffer veto}
\begin{figure}[htb]
    \centering
    \includegraphics[width=\linewidth]{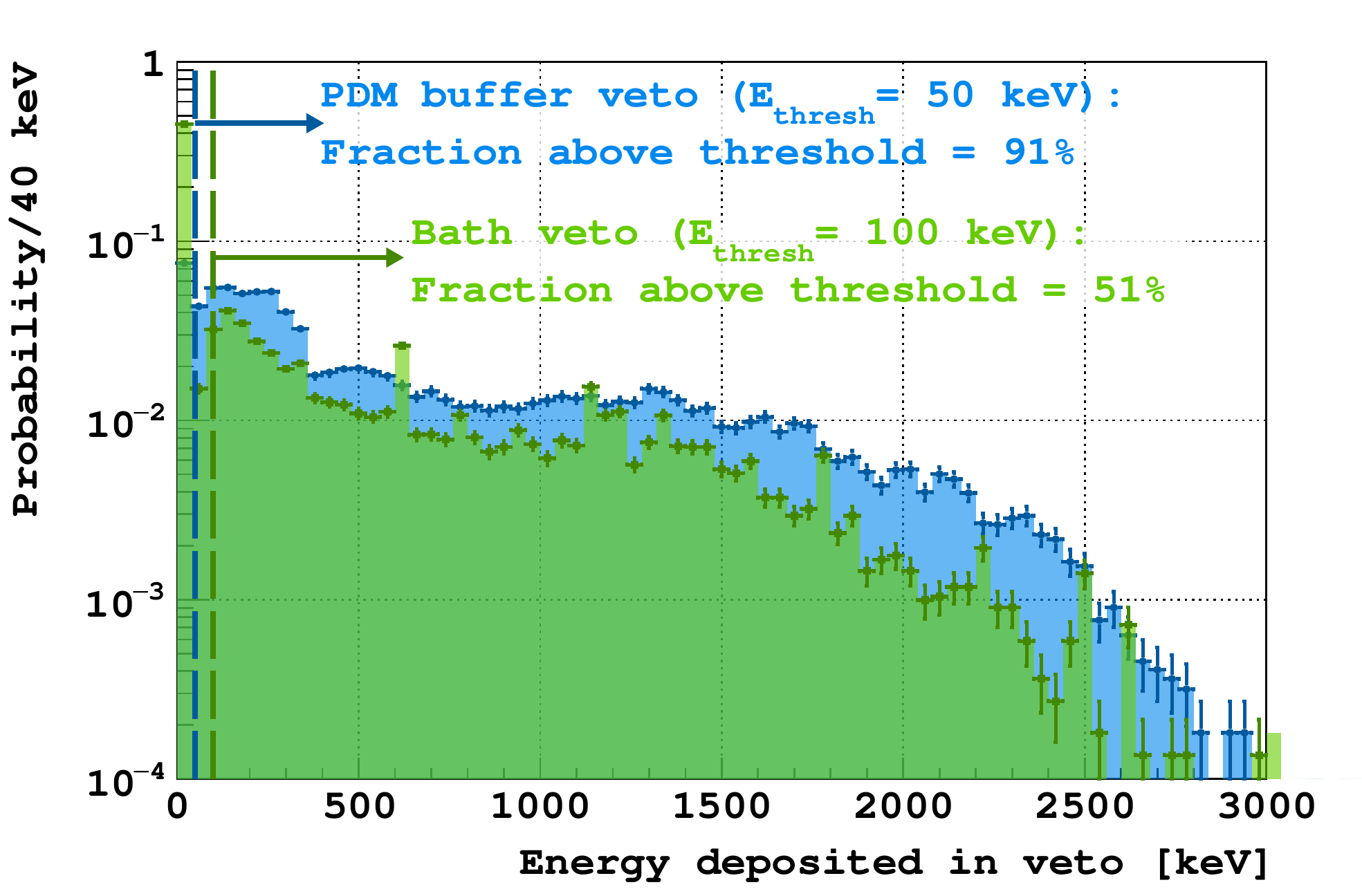}
    \caption{Energy deposited in (blue) the \PDM\ buffer veto and (green) bath veto for simulated \grs\ from the photoelectronics with \SI{<3}{\keV} single-scatters in the \TPC\ fiducial volume, considering both vetoes independently.}
    \label{fig:veto_edep}
\end{figure}
The dominant \gr\ background source is the photoelectronics.
Since fiducialization along the vertical axis is not possible, low-energy \xrs\ and \grs\ that preferentially scatter in the first \SI{10}{\cm} of \LAr\ are not mitigated by fiducial cuts.
Instead, the \PDM\ buffer shields the \TPC\ from such backgrounds while still allowing those that scatter in it to be tagged.

Figure~\ref{fig:veto_edep} shows the energy deposited in the vetoes for simulated \grs\ originating in the photoelectronics that produced single-scatters below \SI{3}{\keV} in the fiducial volume. 
Simulations indicate that the buffer veto can achieve a light yield \SI{\geq4}{\pe\per\keV}, making a \DSLMPDMBufferVetoThreshold\ threshold realistic, and that lower thresholds only marginally improve their efficiency.
Since \grs\ can be absorbed in inactive materials after scattering in the \TPC, only \SI{51}{\percent} are tagged by the bath veto.
However, since they must pass through the \PDM\ buffer veto before reaching the \TPC, \SI{91}{\percent} of the \grs\ that penetrate the buffer and produce a background event are tagged by it.

While the \PDM\ buffer veto and \TPC\ share instrumentation, optical simulations show that pulse shape discrimination efficiently separates scintillation in the buffers from \STwo, and the fraction of light concentrated in the top or bottom \PDM\ array allows these signals to be distinguished from \SOne.

\subsection{$\beta$-decay backgrounds}\label{ssec:backgrounds:beta}

\begin{figure}[htb]
    \centering
    \includegraphics[width=\linewidth]{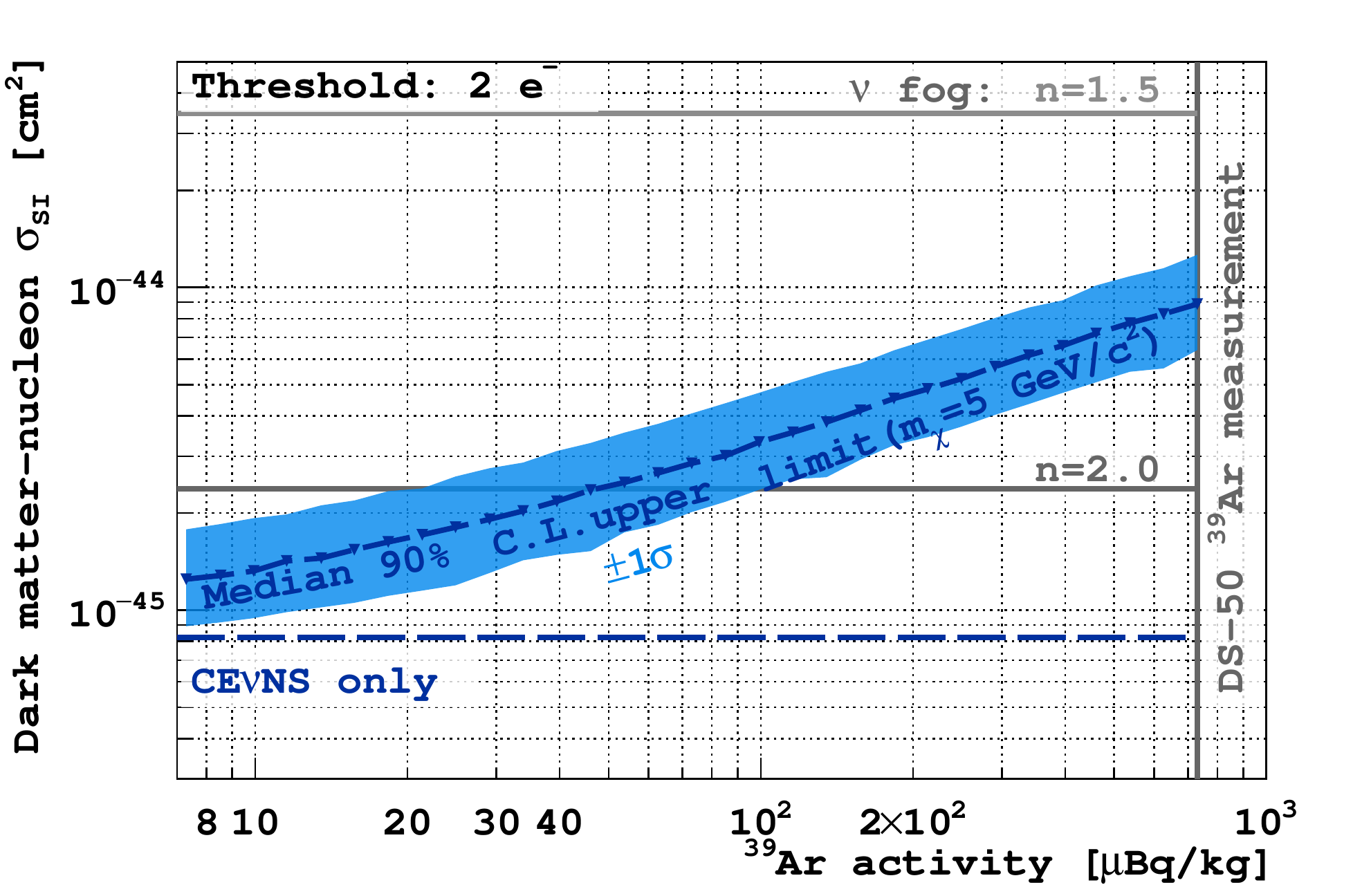}
    \caption{Median \NinetyPerCentCL\ upper limits and $1\sigma$ expectation band on \DSLowMassMidMass\ \DM\ at varying \isotope{39}{Ar} activity.}
    \label{fig:ar39_sensitivity}
\end{figure}

Two naturally-present \bta-emitters have been observed in \UAr: \isotope{39}{Ar}, which \DSf\ measured with a specific activity of \DSfUArArThreeNineActivity, and \isotope{85}{Kr}, at \DSfDdKrEightFiveActivity~\cite{darkside_collaboration_darkside-50_2018}.
Improvements to the \UAr\ extraction facility~\cite{renshawProcuring50Tonnes2018,boulayDEAP3600DiscussionMultiHundred2018} are expected to completely remove \isotope{85}{Kr} and significantly reduce the \isotope{39}{Ar} content relative to \DSf's measurement.

Residual \isotope{39}{Ar} can be further suppressed using the \Aria\ facility~\cite{darkside_collaboration_separating_2021}, which will be capable of depleting \isotope{39}{Ar} by a factor of \AriaDepletionPerPass\ at a \AriaSeruciOneRate\ throughput.

Starting with an \isotope{39}{Ar} activity comparable to \DSf's measurement, the \TPC\ can achieve an activity of \DSfUArArThreeNineActivityOverTen\ with one pass through \Aria.
With improved \UAr\ extraction and a second pass, this activity may be brought as low as \DSfUArArThreeNineActivityOverOneHundred.

Potential internal radioisotope activities are summarized in \reftab{tab:internal_activity_scenarios}.
The effects of varying the \isotope{39}{Ar} activity on the sensitivity to \SI{5}{\GeV\per\square\c} \DM\ with a \SI{2}{\el} threshold are illustrated in \reffig{fig:ar39_sensitivity}, assuming \gr\ and neutrino backgrounds as discussed above. 
Due to their similar $\beta$ spectra, if \ce{^85Kr} is present after purification, \reffig{fig:ar39_sensitivity} can be interpreted as the total activity of \ce{^85Kr} and \ce{^39Ar}.

\begin{table}[htb]
    \centering
    \caption{Internal radioisotope activities explored here, with \DSf\ measurements for reference~\cite{darkside_collaboration_darkside-50_2018}.}
    \begin{tabular}{l|cc}\hline\hline  \rule{0pt}{2.5ex}
        & \isotope{85}{Kr} & \isotope{39}{Ar}  \\
        & \multicolumn{2}{c}{[\si{\micro\becquerel\per\kg}]}\\\hline
         \DSf\ & \DSfDdKrEightFiveActivityNumMicro & \DSfUArArThreeNineActivityValMicro \\
         DS-LM & \num{0} & \DSfUArArThreeNineActivityOverOneHundredVal--\DSfUArArThreeNineActivityOverTenVal \\
        \hline\hline 
    \end{tabular}
    \label{tab:internal_activity_scenarios}
\end{table}

\subsection{Cosmogenic backgrounds}
Cosmic-rays create backgrounds by activating detector materials in transit and by producing prompt muon-induced signals during operations.
\FLUKA~\cite{battistoniOverviewFLUKACode2015} simulations of muon-induced showers at \LNGSfullname\ (\LNGS) based on \refcite{emplFlukaStudyUnderground2014} indicate that they pose a negligible background at comparable or greater depths, such as at \BUL\ or \SNOLAB.
Therefore, these backgrounds are not considered further.

The dominant cosmogenic backgrounds are from \UAr\ activation. 
Calculations are performed assuming the cosmic-ray neutron flux parameterized in \refcite{gordon_measurement_2004}; correction factors for different altitudes and locations are obtained following \refcite{ziegler_terrestrial_1998}.
Production rates and cross sections are taken from measurements and calculations in \refscite{saldanha_cosmogenic_2019,amare_cosmogenic_2018} and \EXFOR~\cite{zerkin_experimental_2018} whenever available. 
Otherwise, cross sections are from the \JENDL~\cite{shibata_jendl-4.0:_2011}, \TENDL~\cite{koning_tendl_2019}, or \HEAD~\cite{korovin_high_2010} libraries or computed from the \COSMO~\cite{martoff_cosmo_1992}, \YIELDX~\cite{silberberg_updated_1998}, and \ACTIVIA~\cite{back_activia_2008} codes.

\begin{table}[htb]
    \centering
    \caption{Expected cosmogenically activated isotopes in \UAr\ after shipping from \Urania\ to \Aria, following surface exposure at \Aria, and after shipping from \Aria\ to \LNGS\ or North America.}
    \begin{tabular}{l|ccc}\hline\hline \rule{0pt}{2.5ex}
        & \isotope{39}{Ar} & \isotope{37}{Ar} & \isotope{3}{H}   \\
        & \multicolumn{3}{c}{[\si{\micro\becquerel\per\kg}]}     \\\hline 
    Urania$\rightarrow$Aria & \num{14.7\pm1.3} & \num{806\pm73}& \num{58\pm12}      \\
    Aria (1\,mo., surface) & \num{2.57\pm0.33}& \num{294\pm39}  & \num{9.0\pm2.8}  \\
    Aria$\rightarrow$LNGS   & \num{0.86\pm0.11}& \num{118\pm15}  & \num{3.00\pm0.95}\\
    Aria$\rightarrow$N.\,America&\num{5.73\pm0.73}&\num{483\pm64}& \num{20.0\pm6.3} \\
        \hline\hline 
    \end{tabular}
    \label{tab:uar_activation}
\end{table}

\reftab{tab:uar_activation} shows \isotope{39}{Ar}, \isotope{37}{Ar}, and \isotope{3}{H} yields for ship transport from the \UAr\ extraction site at \Urania\ (Colorado) to \Aria\ (Sardinia), from \Aria\ to \LNGS\ or North America, and per month outside \Aria's underground column.
Atmospheric argon has \AArArFourTwoActivity\ \ce{^42Ar}~\cite{ajajElectromagneticBackgroundsPotassium422019}, likely orders of magnitude lower in \UAr.
At sea level, it is activated by successive neutron captures on \ce{^40Ar} and \ce{^41Ar} and by $\ce{^40Ar}(\alpha,2p)\ce{^42Ar}$ (\SI{14}{\MeV} threshold), at a rate \SI{e6}{\times} lower than \ce{^39Ar}~\cite{zhangEvaluationCosmogenicProduction2022}.
Other isotopes have short half-lives or will be removed by purification.

At \Aria, \SI{2.57\pm0.33}{\micro\becquerel\per\kg\per month} of \ce{^39Ar} will be activated in \UAr\ stored above ground during distillation.
For long campaigns, these effects can be mitigated by storing \UAr\ underground.
If \DSl\ runs at \LNGS\ or a lab comparably far from \Aria, \SI{0.86\pm0.11}{\micro\becquerel\per\kg} of \ce{^39Ar} will be activated in transit.
At North America, the yield will be \SI{5.73\pm0.73}{\micro\becquerel\per\kg}.
Activated \isotope{3}{H} is separated from argon with SAES Getters~\cite{meikrantzTritiumProcessApplications1995} and will be removed \insitu\ while the \UAr\ recirculates; \isotope{37}{Ar} will decay away with a \SI{35}{day} half-life.
Hence, neither isotope was included in sensitivity projections.

\subsection{Neutrons}
Radiogenic neutrons are produced by \alphan\ reactions from trace  \isotope{238}{U}, \isotope{235}{U}, and \isotope{232}{Th} in detector materials. 
Relevant materials have \alphan\ yields around \numrange{e-6}{e-5}~\cite{westerdale_radiogenic_2017}.
Such neutron backgrounds are therefore expected to be subdominant to those from \grs\ from the same isotopes.
These backgrounds are, therefore, not included in this study.

\subsection{Surface backgrounds}

The \TPC's inner surface area is \DSlSurfaceArea.
During construction, \isotope{222}{Rn} progeny will deposit on surfaces exposed to air~\cite{guiseppe_radon_2010}, accumulating as \isotope{210}{Pb}~\cite{nikezicExposures222RnIts2006}.
Due to its \PbTwoOneZeroHalfLife\ half life, its activity will be suppressed by a factor of \num{2132} relative to the deposited \isotope{222}{Rn} progeny; by cleaning surfaces~\cite{zuzelRemovalLonglived222Rn2012} and assembling the \TPC\ in a radon-scrubbed clean room~\cite{pelczarOnlineRadonMonitor2021}, \isotope{210}{Pb} surface $\beta$ and \xr\ activity can be reduced.
Since fiducialization is only in the horizontal plane, cathode cleanliness is particularly important.

While operating, \isotope{222}{Rn} can emanate from materials and plate-out on walls, causing surface backgrounds from its decay chain, up to \isotope{214}{Po}.
These isotopes are efficiently removed from \LAr\ with charcoal radon-traps~\cite{amaudruz_design_2019,harrisonUseActivatedCharcoal2007} or molecular sieves~\cite{ogawaDevelopmentLowRadioactive2020}; \LAr's cold temperature reduces the radon outgassing~\cite{georgievPartitionCoefficientsDiffusion2019}.

\begin{table}[htb]
    \centering
    \caption{Threshold surface activities of \isotope{222}{Rn} and \isotope{210}{Pb} decay chains needed to contribute \SI{<10}{\percent} of the \gr\ background rate, $\mathcal{A}_{\rm thr}$, compared with the activities reported by \DSf~\cite{darkside_collaboration_darkside-50_2018} and \DEAP~\cite{deap_collaboration_search_2019}.}
    \begin{tabular}{c|ccc}\hline\hline 
        Isotope & $\mathcal{A}_{\rm thr}$     & \DSf              & \DEAP            \\
                        &\multicolumn{3}{c}{[\si{\milli\becquerel\per\square\meter}]}\\\hline
        \isotope{222}{Rn} & \num{6.01\pm0.25} & ---               & \num{<5e-3}      \\
        \isotope{210}{Pb} & \num{2.21\pm0.05} & \num{2.51\pm0.01} & \num{0.26\pm0.02}\\
        \hline\hline 
    \end{tabular}
    \label{tab:surface_bgs}
\end{table}

Surface backgrounds can be controlled through radon-scrubbing and mitigation procedures.
To determine the activity at which they pose a significant background contribution, the \isotope{222}{Rn} and \isotope{210}{Pb} decay chains were simulated within the inner \DSLMSurfaceBkgdDepth\ of the \TPC\ walls, following the same procedure as for \grs.
Upper limits on their activity were then set such that surface backgrounds contribute \SI{<10}{\percent} of the \gr\ background rate from \TPC\ components.

Results from these simulations are given in \reftab{tab:surface_bgs}, compared with surface background rates reported by \DSf~\cite{darkside_collaboration_darkside-50_2018} and \DEAP~\cite{deap_collaboration_search_2019}.
Surface activities obtained by other \LAr\ \DM\ detectors are comparable to or below these limits.
As a result, these backgrounds are not further considered.

\subsection{Spurious electron backgrounds}
\label{subsec:ses}

\begin{figure}[htb]
    \centering
    \includegraphics[width=\linewidth]{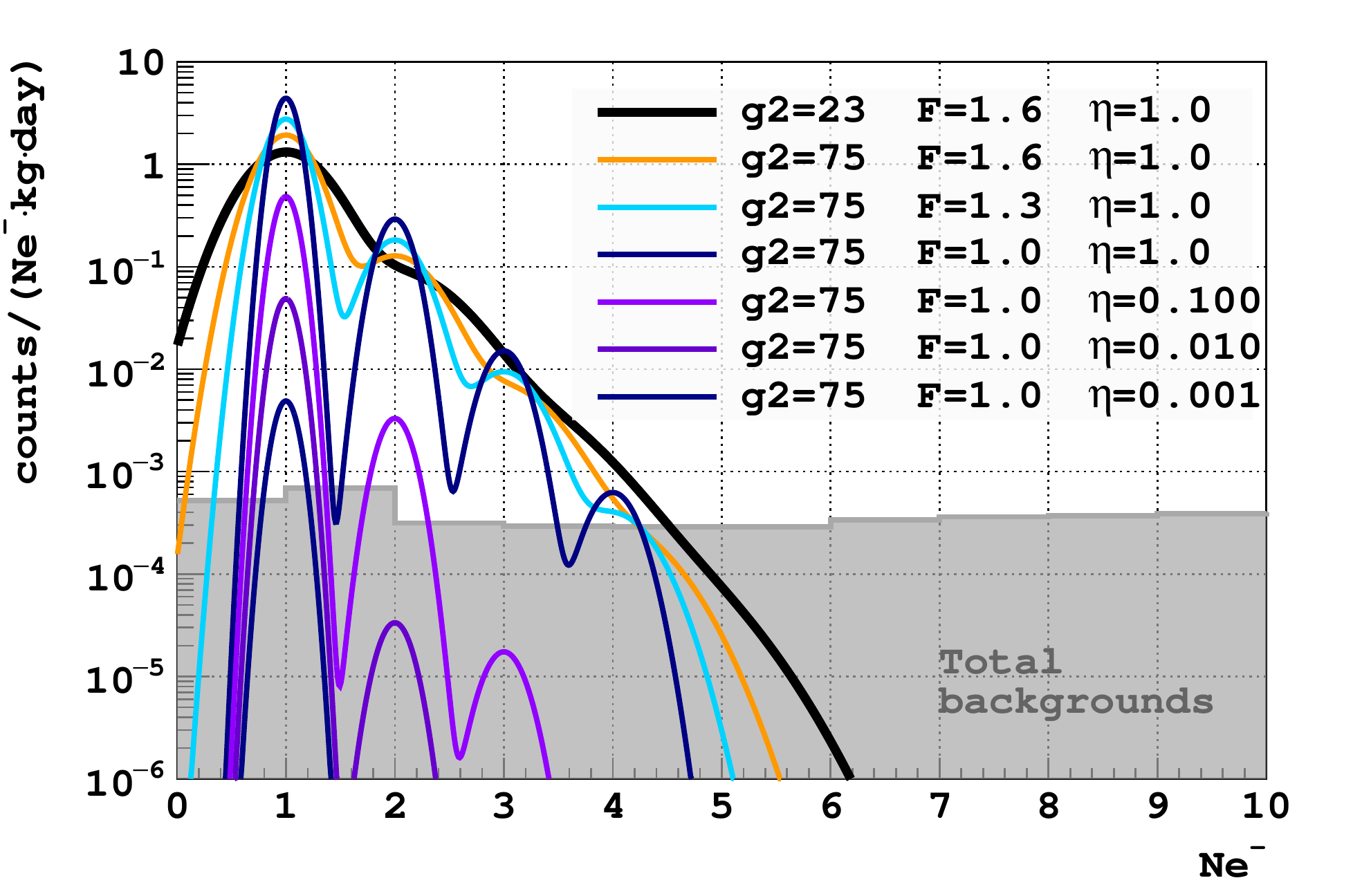}
    \caption{\SE\ spectra scaled from \DSf\ using \refeqn{eq:se_spectra} and \refeqn{eq:se_scaling}, for different electron amplification factors \gTwo, excess noise factor $F$, and impurity scaling factors $\eta$. Backgrounds from other sources are shown for comparison, assuming a \DSfUArArThreeNineActivityOverTen\ \isotope{39}{Ar} activity.}
    \label{fig:se_spectra}
\end{figure}

In \DSf, \SEs\ dominate signals below \SI{4}{\el}.
Leading hypotheses stipulate that they are produced by photo- and electrochemical interactions, rather than by particles scattering in \LAr.
\SEs\ are classified into two categories based on their temporal correlation with preceding progenitor events.
Those within the maximum drift time of electrons in the \TPC, \tdriftmax, are described in \refcite{collaboration_study_2021} and are consistent with photoionization of detector materials.
By requiring the time between pulses to be longer than \tdriftmax, such backgrounds are removed from analysis.

At longer delays, a large fraction of \SEs\ follow preceding \STwo\ signals by a \SI{\sim5}{\milli\second} or \SI{\sim50}{\milli\second} exponential lifetime, with matching horizontal positions; a third component extends to several seconds.
The \SE\ rate is correlated with the total event rate and progenitors' drift time, and it increases when the getter used for purification is turned off.
While a full understanding of \SEs\ requires further investigation, their properties are consistent with impurities capturing and later releasing drifting electrons.
Similar mechanisms have been proposed in xenon~\cite{sorensenTwoDistinctComponents2018}.
In this case, \SEs\ may be reduced with purer \LAr, achievable with \Aria\ and improved \insitu\ purification.
The cold temperature of the \LAr\ bath may also slow impurity diffusion.
Studies of electron attachment in \LAr\ indicate that attachment coefficients can be decreased by tuning the drift field strength~\cite{bakale_effect_1976,swan_electron_1963}.

With improved event reconstruction, it may also be possible to mitigate \SEs\ through their correlations with progenitors. 
After correcting for pulse-finding efficiency, the \Ne\ distribution of \SEs\ in \DSf\ is consistent with a Poisson distribution, implying that \SEs\ above \SI{1}{\el} may be due to pileup.
This explanation is supported by a pulse shape analysis.
Therefore, improved \SE\ reconstruction with higher \Ne\ resolution may allow pileup to be tagged, suppressing backgrounds above \SI{1}{\el}.

Due to their uncertainties, a full \exsitu\ \SE\ model is not possible. 
For most present studies, the \Ne\ value below which they dominate sets an analysis threshold, with \SI{2}{\el} and \SI{4}{\el} thresholds considered.

\begin{figure*}[htb]
    \centering
    \includegraphics[width=0.49\linewidth]{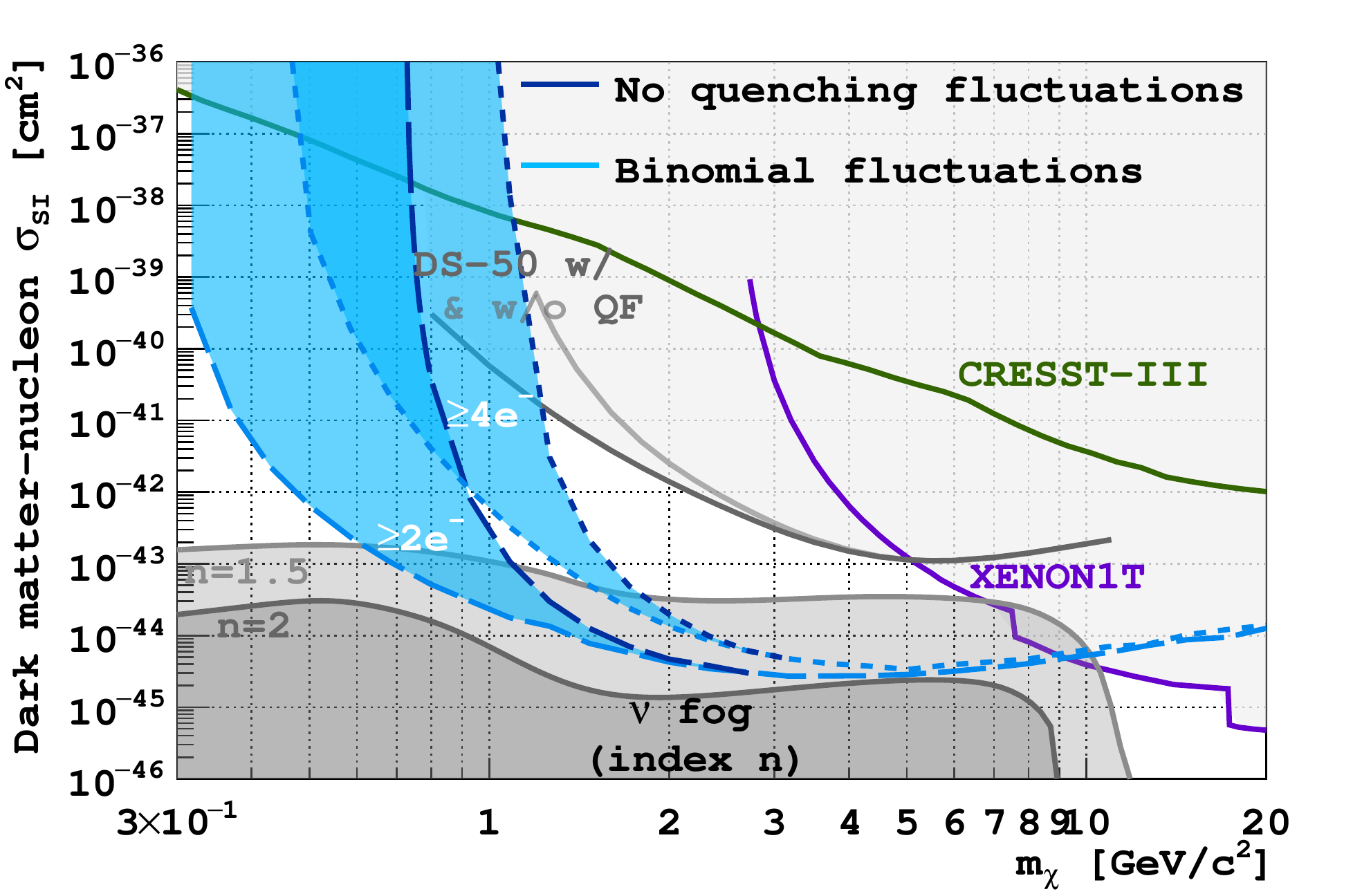}
    \includegraphics[width=0.49\linewidth]{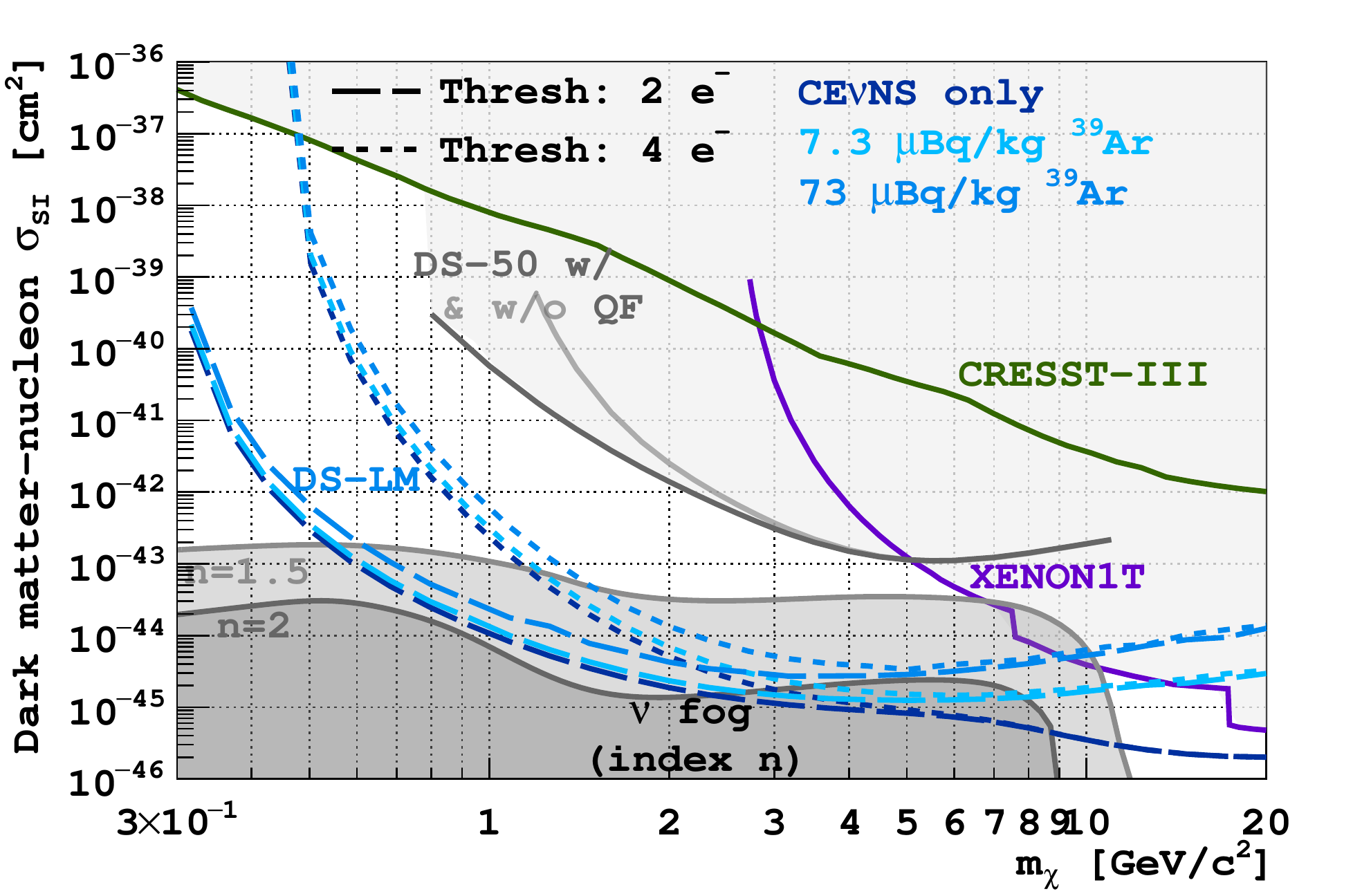}
    \includegraphics[width=0.49\linewidth]{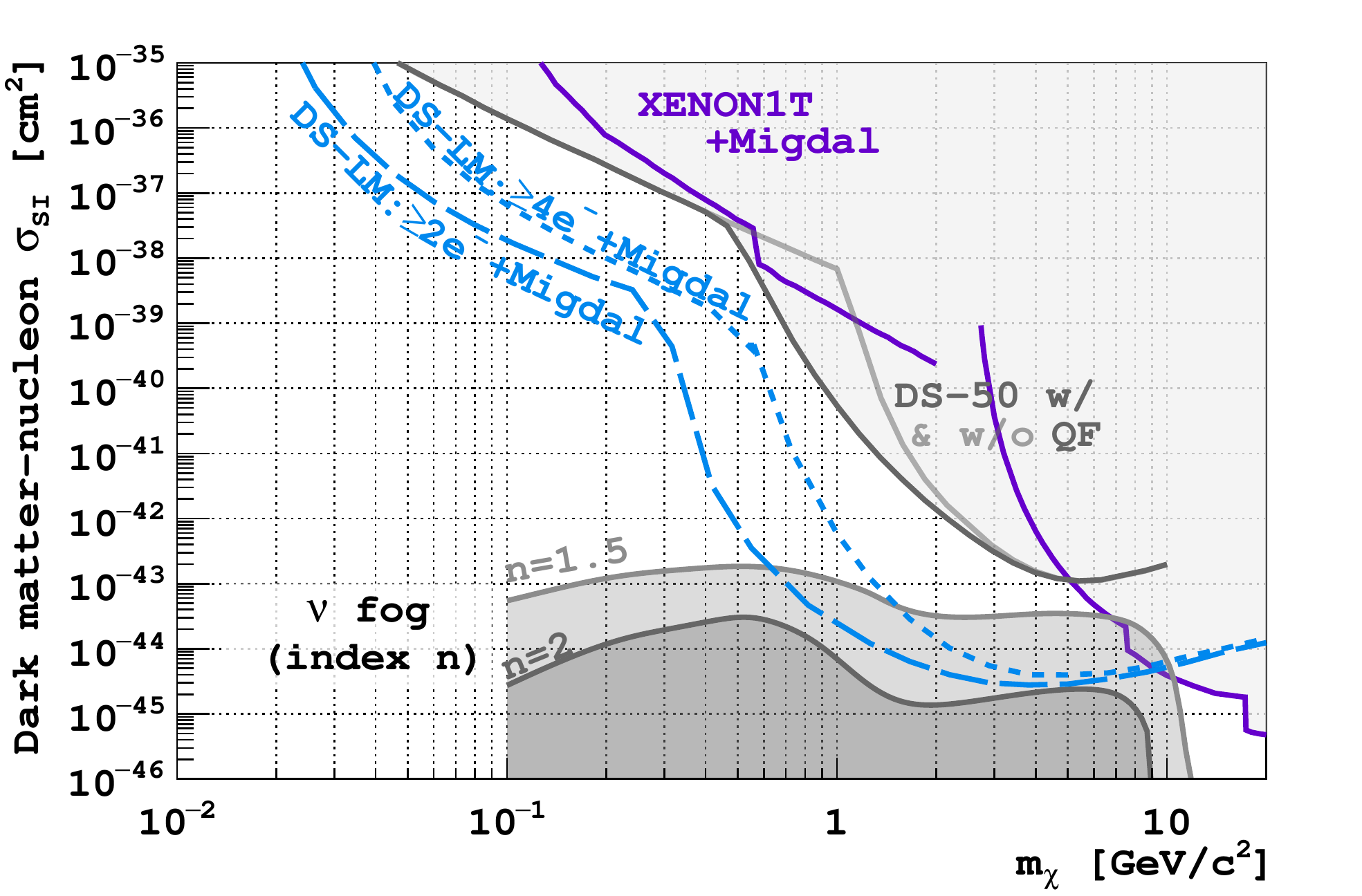}
    \includegraphics[width=0.49\linewidth]{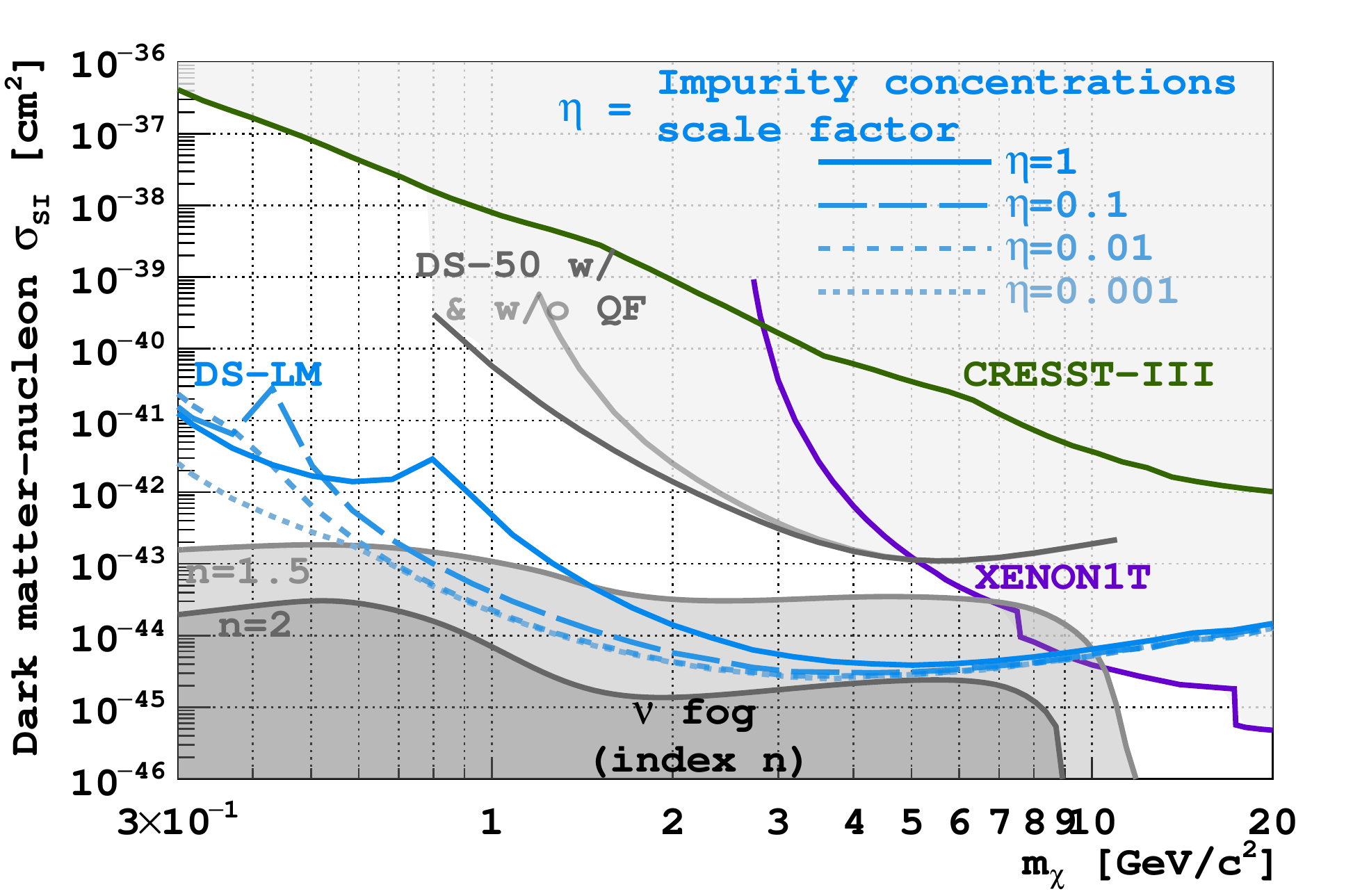}
    \caption{Projected \NinetyPerCentCL\ upper limits on the spin-independent DM-nucleon scattering cross section for \DSLMExposure\ exposure: 
    \emph{(Top, left)} with and without binomial quenching fluctuations. 
    \emph{(Top, right)} with varying thresholds and background rates. 
    \emph{(Bottom, left)} including the Migdal effect. 
    \emph{(Bottom, right)} attempting to model and fit \SE\ backgrounds (see \refeqn{eq:se_spectra}) at varying impurity concentrations relative to \DSf, $\eta$. 
    Unless otherwise stated, projections assume binomial quenching fluctuations and an \ce{^39Ar} activity of \DSfUArArThreeNineActivityOverTen.
    The neutrino fog in \LAr\ with index $n$ representing the resulting impediment to a $3\sigma$ \DM\ observation is shown in shades of gray, calculated to $\text{m}_\chi\SI{=100}{\MeV\per\square\c}$~\cite{ohareNewDefinitionNeutrino2021}.
    Current limits are shown from CRESST-III~\cite{petriccaFirstResultsLowmass2020}, \DSf~\cite{ds50SearchLowmassDark2022,ds50Migdal2022}, and \XENONOT~\cite{aprile_light_2019,aprileSearchLightDark2019}. }
    \label{fig:sensitivity_scenarios}\label{fig:migdal}\label{fig:sefits}
\end{figure*}

To explore effects of \SEs\ beyond their imposition of an \Ne\ threshold, studies will assume a model motivated by \DSf.
This model assumes \SEs\ are produced following an ionization event where some electrons are trapped and later released. 
If $k+1$ electrons are released close in time ($k=0$ corresponding to \SI{1}{\el}), they may appear as a single \STwo\ pulse, leading to an \SE\ with \SI{>1}{\el}. 
Resolution smearing, determined by \gTwo\ and its spatial variance, may cause them to be reconstructed with $n\neq k$ electrons.
The rate of \SEs\ with $n$ electrons is modeled as
\begin{gather}
\begin{aligned}
    \SE(n) =& R\sum_{k=0}\text{P}(k;p)G\left(n;k+1,F\sqrt{\frac{k+1}{\gTwo}}\right) \\ 
    \text{P}(k;p)=& \frac{1}{k!}\left(\frac{p^k}{k+1}-\frac{p^{k+1}}{k+2}\right) ,
\end{aligned}
\label{eq:se_spectra}
\end{gather}
where $R$ is the rate per unit mass;  $G$ estimates the Gaussian probability of reconstructing $n$ electrons, given mean $k$ and standard deviation $F\sqrt{k/\gTwo}$; and $F$ accounts for excess noise beyond \pe\ counting statistics (\emph{e.g.} from spatial \gTwo\ variations). 

$\text{P}(k;p)$ is the probability of $k$ electrons reconstructing in one \STwo, given a pileup probability $p$.
It accounts for the probability of two or more electrons appearing in the same \STwo\ window, which decreases with the exponential decay time of captured electrons.
This model is fit to \DSf\ data and scaled using
\begin{gather}
    \begin{aligned}
      R \propto& R_\text{trig}\times L_\text{drift}^\text{max}\times\eta/M_\text{fid.}  \\
      p \propto& L_\text{drift}^\text{max}\times\eta
    \end{aligned}
    \label{eq:se_scaling}
\end{gather}
where $R_\text{trig}$ is the trigger rate, $L_\text{drift}^\text{max}$ is the maximum drift length, $M_\text{fid}$ is the fiducial mass, and $\eta$ scales the impurity concentration relative to \DSf.

\SE\ spectra for different parameter values are shown in \reffig{fig:se_spectra}.
The bold black curve shows a simple extrapolation from \DSf's best-fit $F$ and \gTwo.
Increasing \gTwo\ and lowering $F$ can decrease the tails of the \SE\ distribution.
Decreasing the impurity concentration by \SIrange{10}{100}{\times} further suppresses \SEs, enabling thresholds as low as \SI{2}{\el}.
Additional suppression of \SEs\ with $\Ne>1$ may be achievable with analysis cuts narrowing the pileup window, thereby decreasing $p$.
Such cuts will be strengthened by improved reconstruction with higher \gTwo\ and lower $F$.

Calculations based on measurements in \refscite{chiouSorptionTransportInert1986,korosObservationsConcerningTemperature1981,crank1968diffusion,liTransportGasesMiscible1993,schmidtchenTemperatureBehaviourPermeation1994,georgievPartitionCoefficientsDiffusion2019}, indicate that poly(methyl methacrylate) (PMMA, or acrylic)  negligibly outgasses impurities in the \SI{87}{\kelvin} \LAr\ bath, and continuous purification during operations will further remove impurities that do enter the system.
It is worth noting that as impurities are reduced, new \SE\ sources may become dominant.

\begin{figure}[htb]
    \includegraphics[width=\linewidth]{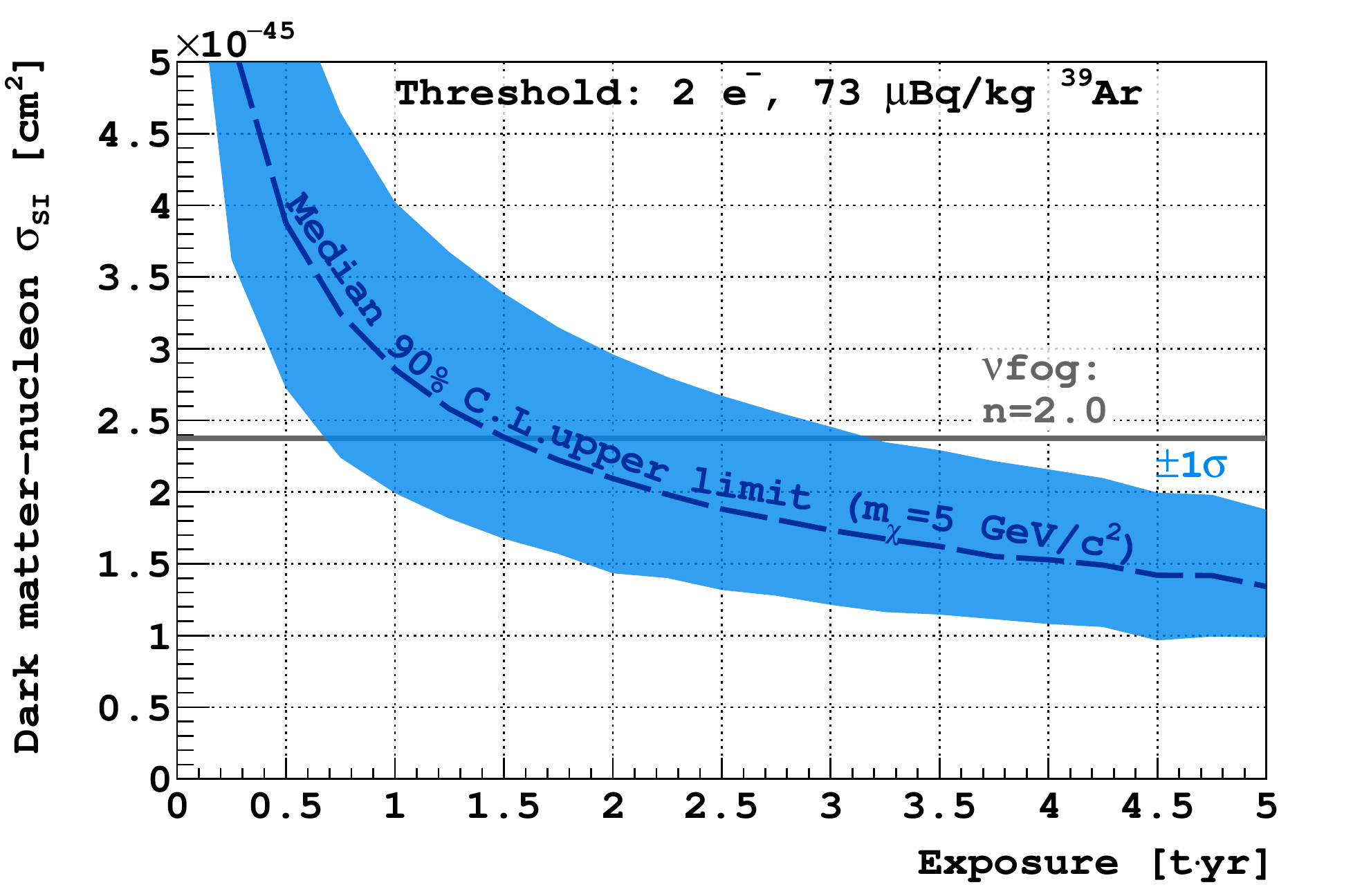}
    \caption{Median \NinetyPerCentCL\ upper limit and $1\sigma$ expectation band for \SI{5}{\GeV\per\square\c} \DM\ at varying exposure.}
    \label{fig:sensitivityVexposure}
\end{figure}

\section{Sensitivity projections}\label{sec:sensitivity}

\begin{figure}[htb]
    \centering
    \includegraphics[width=\linewidth]{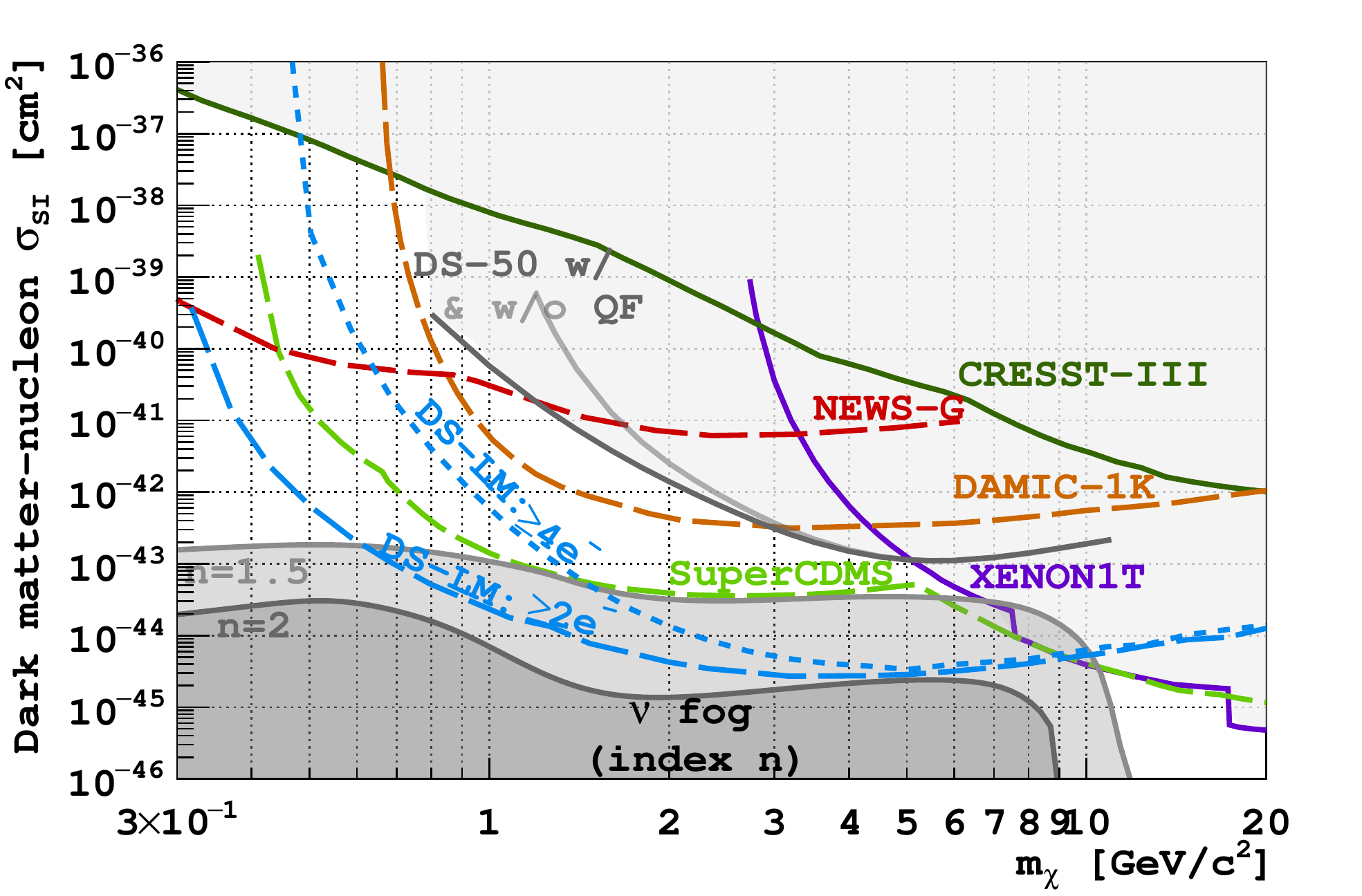}
    \includegraphics[width=\linewidth]{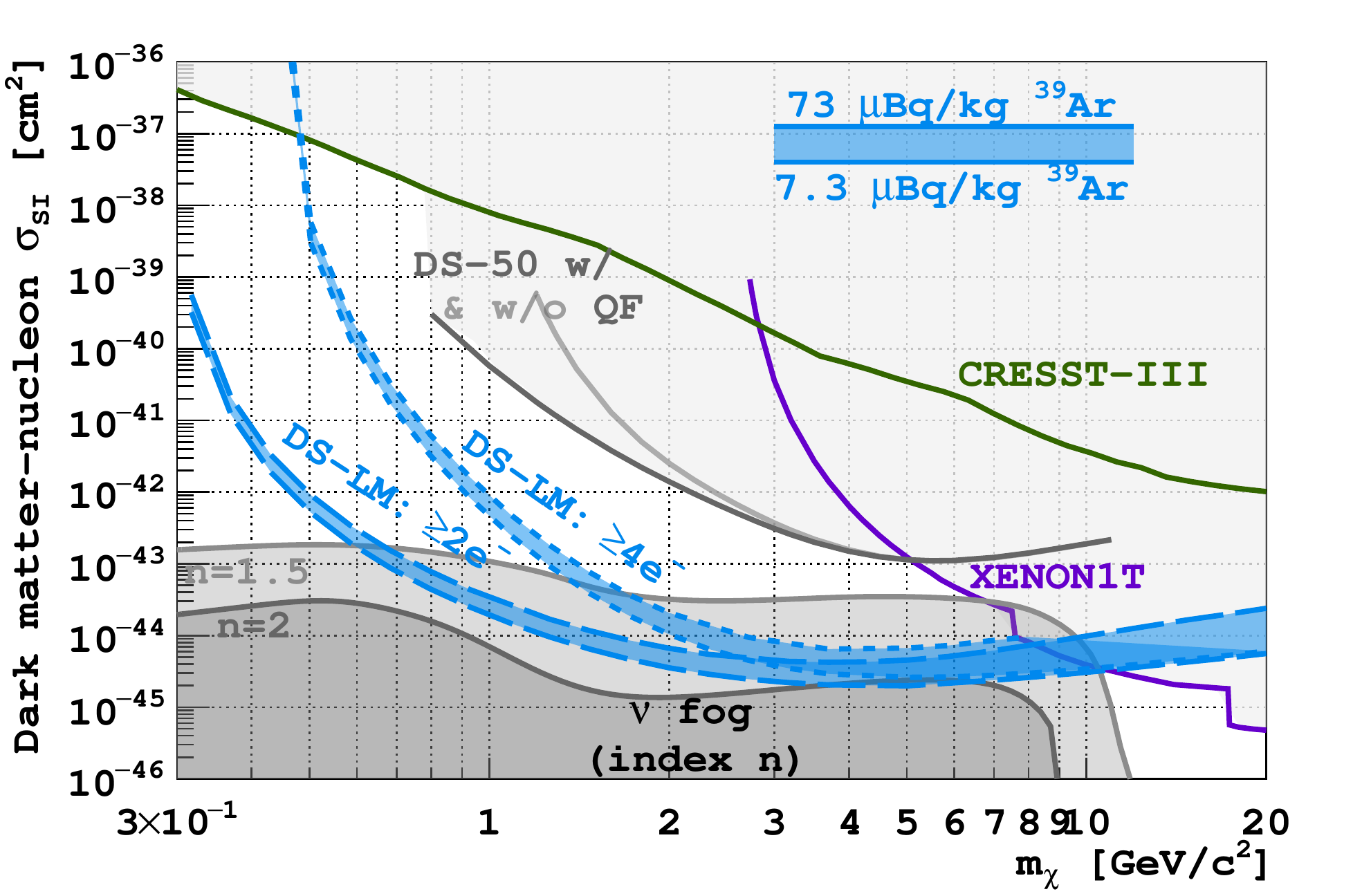}
    \caption{Projected 
    \emph{(Top)} \NinetyPerCentCL\ exclusion curves for the spin-independent \DM-nucleon scattering cross section with \DSfUArArThreeNineActivityOverTen\ of \ce{^39Ar}, compared to (solid) current and (dashed) projected limits.
    \emph{(Bottom)} $3\sigma$ significance evidence contours with a (dashed) \num{2} or (dotted) \SI{4}{\el} threshold and (thick) \DSfUArArThreeNineActivityOverOneHundredVal\ or (thin) \DSfUArArThreeNineActivityOverTen\ of \ce{^39Ar}. 
    Binomial quenching fluctuations and \SI{1}{\tonne\year} exposures are assumed. 
    The neutrino fog in \LAr, with $n$ denoting the impediment to a $3\sigma$ \DM\ observation, is in gray~\cite{ohareNewDefinitionNeutrino2021}.
    Limits from CRESST-III~\cite{petriccaFirstResultsLowmass2020}, \DSf~\cite{ds50SearchLowmassDark2022}, and \XENONOT~\cite{aprile_light_2019} are shown, along with DAMIC-1K~\cite{damiccollaborationConstraintsLightDark2019}, NEWS-G, and SuperCDMS~\cite{supercdmscollaborationProjectedSensitivitySuperCDMS2017} projections.
    }
    \label{fig:sensitivity_projection}
    \label{fig:nscat_discovery}
\end{figure}

\begin{figure*}[htb]
    \centering
    \includegraphics[width=0.49\linewidth]{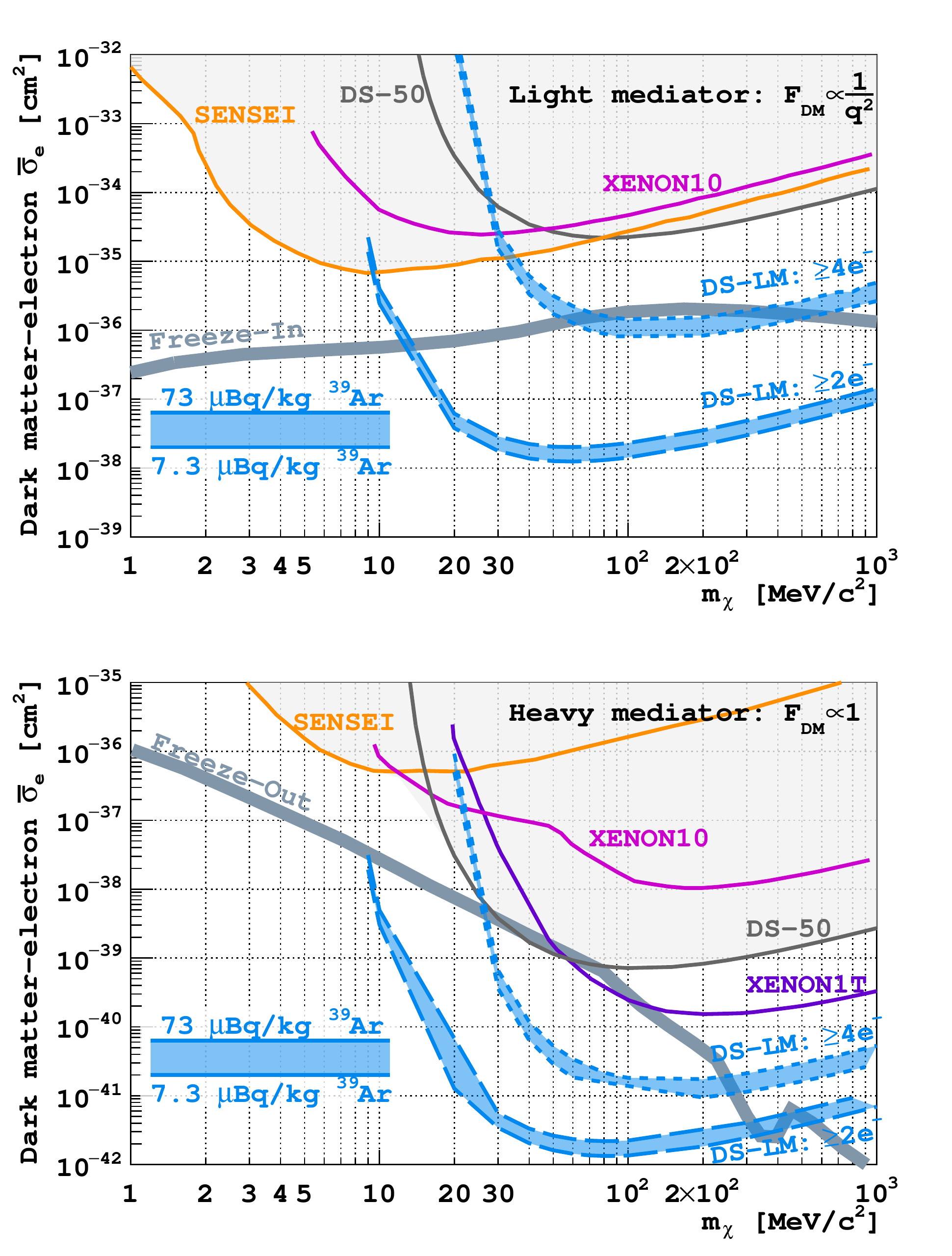}
    \includegraphics[width=0.49\linewidth]{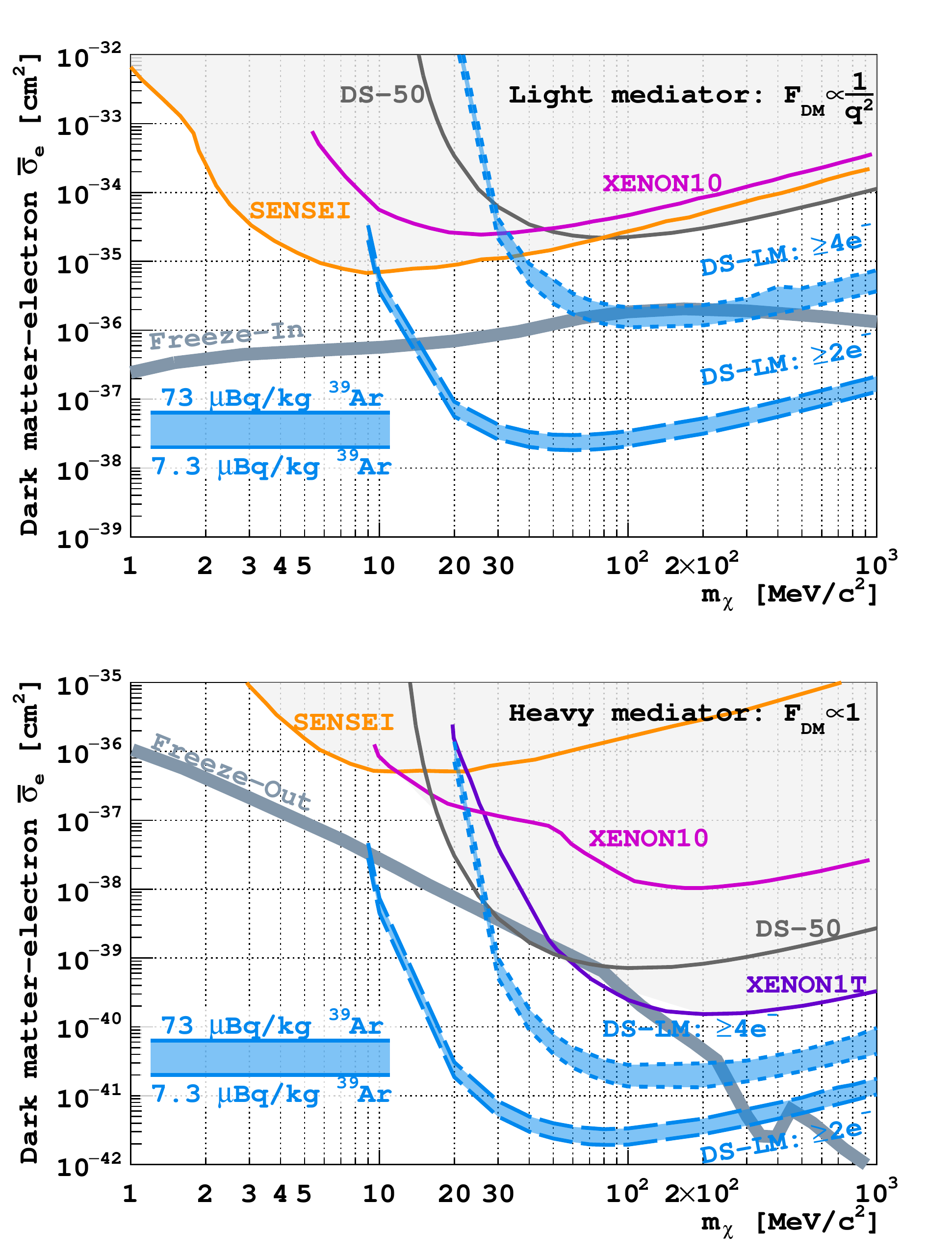}
    \caption{Projected \emph{(left)} \NinetyPerCentCL\ exclusion curves and  \emph{(right)} $3\sigma$ significance evidence contours for \DM-electron couplings with \emph{(top)} light and \emph{(bottom)} heavy mediators.
    Bands show \SI{1}{\tonne\year} contours with \SIrange{7.3}{73}{\micro\becquerel\per\kg} of  \ce{^39Ar}.
    Limits are shown from \DSf~\cite{ds50electronic2022}, SENSEI~\cite{barakSENSEIDirectDetectionResults2020}, XENON10~\cite{essigNewConstraintsProspects2017}, and \XENONOT~\cite{aprile_light_2019}.
    Thick lines show $\bar{\sigma}_e$ giving the relic \DM\ abundance through freeze-in or freeze-out production mechanisms, from \refcite{essigDirectDetectionSubGeV2016}.
    }
    \label{fig:escat_discovery}\label{fig:escat}
\end{figure*}

\DSl's sensitivity is projected for various scenarios using the profile likelihood ratio test statistic (defined in Eq.~11 of \refcite{cowanAsymptoticFormulaeLikelihoodbased2011}) with the $\text{CL}_s$ technique (following \refcite{readPresentationSearchResults2002}) and a Neymann construction to predict median \NinetyPerCentCL\ upper limits for a \DSLMExposure\ exposure.
These tests used the asymptotic approach with an Asimov dataset, as described in \refcite{cowanAsymptoticFormulaeLikelihoodbased2011}, after confirming that it yields indistinguishable results from generating test statistic distributions with a toy Monte Carlo.
Calculations follow the recommendations in \refcite{baxter_recommended_2021}, including the Standard Halo Model described in \refscite{lewinReviewMathematicsNumerical1996,smithRAVESurveyConstraining2007,mccabeEarthTextquotesinglesVelocity2014,schonrichLocalKinematicsLocal2010,bland-hawthornGalaxyContextStructural2016,abuterImprovedGRAVITYAstrometric2021}.

Except where stated otherwise, all projections use binomial quenching fluctuations, the impact of which is illustrated in the top left panel of Fig.~\ref{fig:sensitivity_scenarios}. These results assume the validity of the screening model introduced by Ziegler~\etal~\cite{ziegler_stopping_1985}. 
As shown in Ref.~\cite{the_darkside_collaboration_calibration_2021}, other models give higher ionization yields---up to twice those predicted with the model by Ziegler~\etal---and would therefore predict stronger limits than shown in this work. However, it can not be excluded that a new theoretical screening model could result in weaker projected limits at the lowest energies explored. A dedicated measurement campaign is needed to resolve this issue.

Figure~\ref{fig:ar39_sensitivity} and the top right of \reffig{fig:sensitivity_scenarios} show how lower background rates improve sensitivity at all masses. 
Conservative \isotope{39}{Ar} and \SE\ background reductions enable exclusion sensitivity into the neutrino fog in a \DSLMExposure\ exposure.
Alternative scenarios with further reductions in \SE\ and \ce{^39Ar} background extend this sensitivity down to \SI{1}{\GeV\per\square\c}.
With \DSfUArArThreeNineActivityOverOneHundred\ of \ce{^39Ar}, doubling the \gr\ background rate weakens limits by \SI{<10}{\percent} (\SI{<20}{\percent}) below \SI{5}{\GeV\per\square\c} (\SI{10}{\GeV\per\square\c}).

Figure~\ref{fig:sensitivityVexposure} shows rapid sensitivity growth going from \num{0.1} to \SI{1}{\tonne\year} exposure and modest gains extending to \SI{2}{\tonne\year}.
Longer exposures marginally improve sensitivity, as neutrino backgrounds limit sensitivity.
These trends strengthen at lower \ce{^39Ar} activity.

The top of \reffig{fig:sensitivity_projection} shows that \DSl\ can reach competitive sensitivity in a \DSLMExposure\ exposure.

\subsection{Improvements with the Migdal effect}

Inelastic atomic effects may cause some scattering \DM\ to transfer additional energy to an electron in the target atom, adding an electronic recoil and an \xr/Auger cascade to the nuclear recoil.
This so-called ``Migdal effect'' allows light \DM\ to make higher-energy signals than is possible for a pure nuclear recoil~\cite{ibeMigdalEffectDark2018}.  Given the pure nuclear recoil rate $dR/dE_\text{NR}$, the inelastic rate for producing nuclear and electronic recoil energies $E_\text{NR}$ and $E_\text{ER}$ is
\begin{equation}
    \frac{d^2R}{dE_\text{nr}dE_\text{ER}} = \frac{dR}{dE_\text{NR}}\frac{1}{2\pi}\sum_{n,\ell}\frac{dp^c_{q_e}(n\ell\rightarrow E_\text{ER})}{dE_\text{ER}}
\end{equation}
where $p^c_{q_e}(n\ell\rightarrow E_\text{ER})$ is the probability of an electron with mass $\text{m}_e$ in the $(n\ell)$ shell being ejected with momentum $q_e=\text{m}_e\sqrt{2E_\text{NR}/\text{m}_N}$ in the nuclear rest frame, with mass $\text{m}_N$.
The total deposited energy is $E_\text{NR}+E_\text{ER}+E_{n\ell}$, where $E_{n\ell}$ ($E_{1s}=\SI{3.2}{\keV}$, $E_{2s}=\SI{0.3}{\keV}$, and $E_{2p}=\SI{0.24}{\eV}$) is the binding energy of shell $(n\ell)$.
Signals are modeled as in \refcite{ds50Migdal2022}, summing \NeNR\ from $E_\text{NR}$ with \NeER\ from $E_\text{ER}+E_{n\ell}$.
This approach is conservative, given the non-linearity of \ChargeYieldER.
Values of $p^c_{q_e}$ for isolated atoms are used for all three shells from \refcite{ibeMigdalEffectDark2018}.
The reduced binding energy and the band structure of the valence shell in \LAr\ are not accounted for, rendering this treatment conservative~\cite{catenaAtomicResponsesGeneral2020}.

Significant sensitivity to \DM\ masses as low as \SI{30}{\MeV\per\square\c} can be achieved by exploiting this effect, as illustrated in the bottom left of \reffig{fig:migdal}.
Other effects may give comparable reach~\cite{guoCanSubGeVDark2021a}.

\subsection{Spurious electron background fits}

If R\&D enables \SE\ models, they can be included in the profile likelihood ratio calculation, and the analysis threshold can be lowered, recovering sensitivity.
The effects of such an analysis are explored by modeling \SEs\ with \refeqn{eq:se_spectra}, with $F=1$, $\gTwo=\DSLMgTwoNominal$, and a total event rate of \SI{0.8}{\hertz}, as estimated from simulations.
The effects of varying $\eta$, the impurity concentration relative to \DSf, are explored.

The results of these fits with a \SI{2}{\el} threshold are shown in the bottom right of \reffig{fig:sefits}.
Modeling \SEs\ may extend sensitivity down to \SI{200}{\MeV\per\square\c} masses.
Kinks in the projected exclusion curves are due to \DM\ spectra that closely match the \SE\ spectrum predicted by a given $\eta$.

\subsection{Discovery sensitivity}

The bottom of \reffig{fig:nscat_discovery} shows \DSl's potential for observing evidence of \DM\ at $3\sigma$ significance with varying \isotope{39}{Ar} concentrations and analysis thresholds.
In \SI{1}{\tonne\year}, a \SI{4}{\el} threshold can reach the $n=1.5$ neutrino fog above \SI{1.7}{\GeV\per\square\c}, with significant sensitivity down to \SI{0.5}{\GeV\per\square\c}.
A \SI{2}{\el} threshold extends the reach to \SI{0.3}{\GeV\per\square\c}, with masses above \SI{0.7}{\GeV\per\square\c} within the fog. 
Decreasing the \isotope{39}{Ar} activity improves sensitivity at all masses.

\begin{table}[htb]
    \centering
    \caption{\DM\ masses above which evidence (discovery) contours are within the $n=1.5$ solar neutrino fog at $3\sigma$  ($5\sigma$) significance, up to \SI{\sim10}{\GeV\per\square\c}.}
    \begin{tabular}{cc|cc}\hline\hline\rule{0pt}{2.5ex}
        \Ne\ threshold & \ce{^39Ar} activity & $3\sigma$ & $5\sigma$ \\
        \si{[\el]} & \si{[\micro\becquerel\per\kg]} & \multicolumn{2}{c}{\si{[\GeV\per\square\c]}} \\\hline
        2 & \DSfUArArThreeNineActivityOverOneHundredVal & \num{0.60} & \num{0.68} \\ 
        2 & \DSfUArArThreeNineActivityOverTenVal & \num{0.68} & \num{0.79} \\
        4 & \DSfUArArThreeNineActivityOverOneHundredVal & \num{1.42} & \num{1.67} \\
        4 & \DSfUArArThreeNineActivityOverTenVal & \num{1.71} & \num{2.12} \\
        \hline\hline
    \end{tabular}
    \label{tab:nufog_disc_masses}
\end{table}

An observation rejecting the background-only hypothesis at $3\sigma$ significance would constitute evidence for \DM, while $5\sigma$ amounts to a discovery.
\reftab{tab:nufog_disc_masses} summarizes the masses for which $3\sigma$ and $5\sigma$ significance is reached within the $n=1.5$ neutrino fog.
Evidence for \DM\ would warrant follow-up studies to either confirm or refute the possible signal and to test if it can be explained by a poorly-understood background like \SEs.
These tests could include searching for annual modulation in the excess and searches with a liquid xenon \TPC, where the \SEs\ behave differently than in \LAr, or with entirely different technology with different low-energy systematics, like SBC~\cite{alfonso-pitaSnowmass2021Scintillating2022} or SuperCDMS~\cite{supercdmscollaborationProjectedSensitivitySuperCDMS2017}, among others.
In order to detect compelling evidence for \DM, it is critical to better understand SEs.

\subsection{Electron-scattering dark matter}

\DSl\ will be sensitive to \DM\ with electronic couplings, via a vector mediator with mass \DarkPhotonMassSymbol.
As in \refcite{the_darkside_collaboration_constraints_2018}, limiting cases of \mbox{$\DarkPhotonMassSymbol\gg 1/a_0$} (heavy mediator) and \mbox{$\DarkPhotonMassSymbol\ll 1/a_0$} (light mediator) are considered, giving \DM\ form factors $F_\mathrm{DM}(q)$ of $1$ or $1/(a_0 q)^2$, where $a_0$ is the Bohr radius and $q$ is the momentum transfer.
\reffig{fig:escat} shows the projected \NinetyPerCentCL\ exclusion curves and $3\sigma$ evidence contours with \SI{1}{\tonne\year} exposure.
Sensitivity to heavy (light) mediators with cross sections down to \SI{e-42}{\square\cm} (\SI{e-38}{\square\cm}) may be reached at \SI{100}{\MeV\per\square\c}.

\begin{table}[htb]
    \centering
    \caption{\DM\ masses where \DM\ produced by freeze-in ($\DarkPhotonMassSymbol\ll1/a_0$) or freeze-out ($\DarkPhotonMassSymbol\gg1/a_0$) may be observed at $3\sigma$ (evidence) and $5\sigma$ (discovery) significance.}
    \begin{tabular}{cc|cc|cc}\hline\hline\rule{0pt}{2.5ex}
        \Ne & \ce{^39Ar} & \multicolumn{2}{c|}{$\DarkPhotonMassSymbol\ll 1/a_0$} & \multicolumn{2}{c}{$\DarkPhotonMassSymbol\gg 1/a_0$} \\
        thresh. & activity & $3\sigma$ & $5\sigma$ & $3\sigma$ & $5\sigma$ \\
        \si{[\el]} & \si{[\micro\becquerel\per\kg]} & \multicolumn{2}{c|}{\si{[\MeV\per\square\c]}} & \multicolumn{2}{c}{\si{[\MeV\per\square\c]}} \\\hline
        2 & \DSfUArArThreeNineActivityOverOneHundredVal & \numrange{13}{1000} & \numrange{15}{1000} & \numrange{9}{317} & \numrange{9}{293} \\ 
        2 & \DSfUArArThreeNineActivityOverTenVal & \numrange{15}{1000} & \numrange{16}{1000} & \numrange{9}{291} & \numrange{10}{270}  \\
        4 & \DSfUArArThreeNineActivityOverOneHundredVal & \numrange{66}{404} & ---  & \numrange{27}{256} & \numrange{27}{236} \\
        4 & \DSfUArArThreeNineActivityOverTenVal & --- & --- & \numrange{28}{230} & \numrange{29}{192}\\
        \hline\hline
    \end{tabular}
    \label{tab:erdm_disc_masses}
\end{table}

\DM\ coupled to electrons via a dark photon with $\DarkPhotonAlphaSymbol\equiv g_D^2/4\pi$, where $g_D$ is the $U(1)_D$ gauge coupling, can be produced at the relic abundance through the freeze-in mechanism if \mbox{$\DarkPhotonMassSymbol\ll 1/a_0$} and the freeze-out mechanism if \mbox{$\DarkPhotonMassSymbol\gg 1/a_0$}~\cite{essigDirectDetectionSubGeV2016}.
\reffig{fig:escat_discovery} shows the \DM-electron scattering cross section $\bar{\sigma}_e$ that gives the relic abundance for \DM\ of mass \WIMPMassSymbol\ with $\DarkPhotonAlphaSymbol=0.5$ and either $\DarkPhotonMassSymbol\rightarrow0$ or  $\DarkPhotonMassSymbol=3\WIMPMassSymbol$ for light and heavy mediators, respectively.
Away from resonances such as $\DarkPhotonMassSymbol=2\WIMPMassSymbol$, these curves vary little with choice of \DarkPhotonMassSymbol\ and \DarkPhotonAlphaSymbol~\cite{essigDirectDetectionSubGeV2016}.
\reftab{tab:erdm_disc_masses} summarizes \WIMPMassSymbol\ ranges for which \DSl\ may be able to observe \DM\ with $\bar{\sigma}_e$ predicted by either mechanism with at least $3\sigma$ or $5\sigma$ significance.

\subsection{Solar neutrino sensitivity}\label{ssec:projections:solar}

\CEnNS\ from solar neutrinos presents an opportunity to study solar neutrinos through a flavor-universal channel.
This reaction was first detected by COHERENT~\cite{COHERENT:2017ipa,COHERENT:2020iec}, enabling such studies.
With a \SI{2}{\el} (\SI{4}{\el}) threshold, an \ce{^39Ar} activity of \SI{14.6}{\micro\becquerel\per\kg} (\DSfUArArThreeNineActivityOverOneHundred) is required to detect solar neutrinos with $5\sigma$ significance in \SI{1}{\tonne\year}.

\section{Ideas for further improvements and upgrades}\label{sec:improvements}
The small size and relaxed light yield requirements afford \DSl\ the flexibility to improve its sensitivity through design features, beyond those in the conceptual design discussed in this paper, either as improvements to the baseline design or as future upgrades, pending additional R\&D.

\DSf\ found that \SEs\ may largely be due to drifting electrons capturing on impurities and later being released. 
Improvements in the purification system targeting these impurities or modifications that avoid their introduction may reduce \SEs, as may techniques for tagging piled-up \SEs\ or fitting them in data.
They may also be reduced by shortening the \TPC\ while maintaining the same target mass or by decreasing the total event rate in the fiducial volume.
\TPB\ may be one impurity responsible for \SEs: it is soluble in \LAr\ (possibly at the $\mathcal{O}(\si{\ppb})$-level)~\cite{asaadiEmanationBulkFluorescence2019} and has $\mathcal{O}(\SI{1}{\milli\second})$ excited states observed in its scintillation~\cite{stanfordSurfaceBackgroundSuppression2018a}.
Alternatives like the PEN wavelength shifter, VUV-sensitive \SiPMs~\cite{igarashiPerformanceVUVsensitiveMPPC2016,pershingPerformanceHamamatsuVUV42022a}, or Xe-doping~\cite{galbiatiPulseShapeStudy2021,voglScintillationOpticalProperties2022} may therefore reduce \SEs.

Doping \LAr\ may also extend sensitivity to lower \DM\ masses~\cite{laverneScintillationIonizationAllenedoped1996,kubotaIonizationYieldXenondoped1974}: additives with lower ionization energies can increase the yield and lower the energy threshold~\cite{ichinoseEnergyResolutionMeV1990}.
At higher concentrations, additives with light nuclei---including hydrogenous photo-ionizing dopants~\cite{andersonNewPhotosensitiveDopants1986}---may offer targets with ideal kinematic coupling to light \DM\ and sensitivity to spin-dependent interactions. 
Doping \LAr\ in a second phase may be akin to a ``beam-on/beam-off'' experiment for \DM\ candidates detectable only by the doped target. 
Since the dominant low-energy backgrounds are \SEs, changing the ionization properties of the \LAr\ with dopants may also disambiguate instrumental noise from \DM\ signals.
\DSl's small size will afford it the flexibility for such upgrades through a phased approach.

\section{Conclusion}\label{sec:conclusions}
These studies show that a tonne-scale dual-phase \LArTPC\ with existing technology can reach sensitivity to \DM\ with nuclear couplings in the solar neutrino fog with a \DSLMExposure\ exposure.
This can be achieved with a detector similar to \DSf, scaled to a larger target mass with available \UAr\ further suppressed in \isotope{39}{Ar} by \Aria. 
In addition to increasing the exposure, the larger mass enables self-shielding, using horizontal fiducialization and the PDM buffer vetoes, to further suppress \gr\ backgrounds.

Present uncertainties in modeling the ionization response of \LAr\ to low-energy nuclear and electronic recoils hinder analyses at lower masses: the top left panel of \reffig{fig:sensitivity_scenarios} illustrates the effects of how
ionization yield fluctuations are modeled, while 
\refcite{the_darkside_collaboration_calibration_2021} shows that the choice
in nuclear recoil screening function may increase \ChargeYieldNR\ by nearly
a factor of two below \SI{10}{\keVr}, relative to the model by Ziegler
\etal~\cite{ziegler_stopping_1985} considered in this work.
New measurements below \SI{10}{\keV}, similar to those in \refcite{scene_collaboration_measurement_2015,ARIS}, may address these uncertainties and benefit \DSl.

Improved radiopurity, including low-radioactivity \SiPMs, and the \gr\ veto system enable a design in which \gr\ backgrounds are subdominant to those from solar neutrinos. 
The strongest factors for improving sensitivity are further removing \isotope{39}{Ar}, with expected gains down to \DSfUArArThreeNineActivityOverOneHundred, and lowering the energy threshold.
The relatively small target mass allows its \UAr\ to be depleted by \Aria\ in a feasible timescale.
Little sensitivity is gained with exposures larger than \SI{\sim1}{\tonne\year}, characteristic of \DM\ searches in the neutrino fog.
Even if SEs are not reduced, the ability to deplete \ce{^39Ar} in Aria will enhance \DSl's sensitivity, and the stronger electroluminescence field will enhance analysis capabilities for the lowest-energy signals. 
While these improvements will extend \DSl's sensitivity, especially at lower masses, this fog is already within reach for readily-realizable scenarios.
More novel upgrades in a second phase of the experiment can mitigate backgrounds to reach into the neutrino fog for a wider range of \DM\ masses, and they can extend sensitivity to lighter candidates.
\DSl's small size and flexible design will allow these upgrades to be made, including possible modifications to enhance the detectors response or decrease SE backgrounds, if more is learned of their causes after the detector is first commissioned.

\section{Acknowledgments}
The DarkSide Collaboration would like to thank LNGS and its staff for invaluable technical and logistical support. 
This report is based upon work supported by the U. S. National Science Foundation (NSF) (Grants No. PHY-0919363, No. PHY-1004054, No. PHY-1004072, No. PHY-1242585, No. PHY-1314483, No. PHY- 1314507, associated collaborative grants, No. PHY-1211308, No. PHY-1314501, No. PHY-1455351 and No. PHY-1606912, as well as Major Research Instrumentation Grant No. MRI-1429544), the Italian Istituto Nazionale di Fisica Nucleare (Grants from Italian Ministero dell’Istruzione, Università, e Ricerca Progetto Premiale 2013 and Commissione Scientific Nazionale II), the Natural Sciences and Engineering Research Council of Canada,
SNOLAB, and the Arthur B. McDonald Canadian Astroparticle Physics Research Institute. 
We acknowledge the financial support by LabEx UnivEarthS (ANR-10-LABX-0023 and ANR18-IDEX-0001), the São Paulo Research Foundation (Grant FAPESP-2017/26238-4), Chinese Academy of Sciences (113111KYSB20210030) and National Natural Science Foundation of China (12020101004).
The authors were also supported by the Spanish Ministry of Science and Innovation (MICINN) through the grant PID2019-109374GB-I00, the ``Atraccion de Talento'' grant 2018-T2/TIC-10494, 
the Polish NCN (Grant No. UMO-2019/33/B/ST2/02884), the Polish Ministry of Science and Higher Education (MNiSW, grant number 6811/IA/SP/2018), the International Research Agenda Programme AstroCeNT (Grant No. MAB/2018/7) funded by the Foundation for Polish Science from the European Regional Development Fund, the European Union’s Horizon 2020 research and innovation program under grant agreement No 952480 (DarkWave), the Science and Technology Facilities Council, part of the United Kingdom Research and Innovation, and The Royal Society (United Kingdom), and IN2P3-COPIN consortium (Grant No. 20-152).  
I.F.M.A is supported in part by Conselho Nacional de Desenvolvimento Científico e Tecnológico (CNPq). We also wish to acknowledge the support from Pacific Northwest National Laboratory, which is operated by Battelle for the U.S. Department of Energy under Contract No. DE-AC05-76RL01830.
This research was supported by the Fermi National Accelerator Laboratory (Fermilab), a U.S. Department of Energy, Office of Science, HEP User Facility. Fermilab is managed by Fermi Research Alliance, LLC (FRA), acting under Contract No. DE-AC02-07CH11359.
For the purpose of open access, the authors have applied a Creative Commons Attribution (CC BY) public copyright license to any Author Accepted Manuscript version arising from this submission.

\bibliographystyle{biblio}
\bibliography{references}
\end{document}